\documentclass[article,oneauthor,pdftex,10pt,a4paper]{mdpi-arxiv}
\Title{Atom-Diffraction from Surfaces with Defects:\\ A Fermatian, Newtonian and Bohmian Joint View}

\Author{\'Angel S. Sanz\orcidA{}}

\AuthorNames{\'Angel S. Sanz}

\address[1]{%
Department of Optics, Faculty of Physical Sciences,
Universidad Complutense de Madrid,\\
Pza.\ Ciencias 1, Ciudad Universitaria, 28040 Madrid, Spain;
a.s.sanz@fis.ucm.es}

\corres{Correspondence: a.s.sanz@fis.ucm.es; Tel.: +34-91-394-4673}

\abstract{Bohmian mechanics, widely known within the field of the quantum foundations, has been a
quite useful resource for computational and interpretive purposes in a wide variety of practical problems.
Here, it is used to establish a comparative analysis at different levels of approximation
in the problem of the diffraction of helium atoms from a substrate consisting of a defect with axial symmetry
on top of a flat surface.
The motivation behind this work is to determine which aspects of one level survive in the next level of
refinement and, therefore, to get a better idea of what we usually denote as quantum-classical correspondence.
To this end, first a quantum treatment of the problem is performed with both an approximated hard-wall model
and then with a realistic interaction potential model.
The interpretation and explanation of the features displayed by the corresponding diffraction intensity patterns
is then revisited with a series of trajectory-based approaches: Fermatian trajectories (optical rays), Newtonian
trajectories and Bohmian trajectories.
As it is seen, while Fermatian and Newtonian trajectories show some similarities, Bohmian trajectories behave
quite differently due to their implicit non-classicality.}

\keyword{atom-surface scattering; bohmian mechanics; matter-wave optics; diffraction; vortical~dynamics}

\newcommand{\bd}{\begin{displaymath}}
\newcommand{\ed}{\end{displaymath}}
\newcommand{\be}{\begin{equation}}
\newcommand{\ee}{\end{equation}}
\newcommand{\ba}{\begin{eqnarray}}
\newcommand{\ea}{\end{eqnarray}}

\begin{document}


\section{Introduction}
\label{sec1}

In the last several decades, there has been a fruitful and beneficial transfer of the ideas involved
in David Bohm's formulation of quantum mechanics \cite{bohm:PR:1952-1,bohm:PR:1952-2,bohm-hiley-bk,holland-bk}
from the domain of the quantum foundations to the arena of the applications
\cite{chattaraj-bk,hughes-bk,oriols-bk,sanz-bk-2,benseny:EPJD:2014}.
The conceptual and mathematical background provided by Bohmian mechanics
\cite{bohm-hiley-bk,holland-bk,duerr-bk:2009,duerr-bk:2013,sanz-bk-1} has become a resourceful
tool to investigate quantum problems from an alternative viewpoint regardless of the always ongoing
hidden-variable debate.
This includes both fresh interpretations to (known and also new) quantum phenomena and
novel implementations of alternative numerical algorithms to tackle them \cite{wyatt-bk}.
The essential ingredients of Bohmian mechanics have also inspired methodologies and descriptions aimed at
providing effective trajectory-like explanations of wave phenomena beyond the quantum realm \cite{sanz:EJP-arxiv:2017}.
For instance, Bohmian-like trajectories have been synthesized from experimental data with light
\cite{kocsis:Science:2011}, while Bohmian-like behaviors have been recreated in classical fluid-dynamics experiments
\cite{couder:Nature:2005,couder:PRL:2006,couder:JFluidMech:2006,couder:PNAS:2010,bush:PNAS:2010,bush:PRE:2013,bush:ARFM:2015}.

Getting back to quantum mechanics, one of the advantages of Bohmian mechanics is, perhaps, its~capability to put on the same level quantum and classical analyses or descriptions of the same
physical phenomenon by virtue of the concept of trajectory, well defined in both contexts.
Now, because Bohmian trajectories are in compliance with quantum mechanics, they can be considered
to be at a descriptive level above classical trajectories.
Thus, an interesting question that naturally arises is how much or what kind of information
is kept when passing from a descriptive level to the next one.
This is precisely the question addressed here.
To this end, a realistic working system is considered, although it is still simple enough to provide
a clear answer to the question.
Specifically, the phenomenon analyzed is the helium-atom diffraction from a carbon monoxide molecule
(CO) adsorbed on a platinum~(111) single-crystal surface.
This is a system that has been extensively studied in the literature both experimentally and theoretically
\cite{lahee:PRL:1986,lahee:JCP:1987,drolshagen:JCP:1987,graham:JCP:1996,yinnon:JCP:1988,lemoine:JCP:1994-1,lemoine:PRL:1998,choi:JCP:1997-1,choi:JCP:1997-2},
even from a Bohmian viewpoint~\cite{sanz:jcp:2004,sanz:prb:2004}.
An appropriate description and knowledge of the CO-Pt(111) interaction is important to the understanding
of the role of Pt as a catalyst of the electrochemical oxidation of the CO, with industrial and
technological applications (of course, an extensive literature on other analogous systems is also available).
In~order to determine such an interaction, one of the experimental techniques employed is the
He-atom diffraction (or scattering) at low energies (typically energies are between 10~meV and 200~meV,
about three orders of magnitude below the range for low-energy electron diffraction).
This is a rather convenient tool to investigate and characterize surfaces at relatively low energies with
neutral probes, which~provides valuable information about the surface electronic distribution without a
damage of the crystal\mbox{---the atoms} remain a few {\AA}ngstroms above the surface, in contrast to low-energy
electron diffraction, where electrons penetrate a few crystal layers, strongly interacting with the crystal
atoms.

As is well-known, when a matter wave is diffracted by a crystal lattice, either by reflection (the case of
He atoms or low-energy electrons) or by transmission (the case of neutron diffraction or high-energy electrons),
the resulting spatial intensity distribution is characterized by a series of maxima along the so-called Bragg
directions.
The lattice structure can be determined from these characteristic patterns by means of a convenient
modeling.
Sometimes, however, these patterns have distortions due to a break of the translational symmetry
typical of a perfect lattice.
This symmetry-breaking can be induced by intrinsic thermal (lattice) atom vibrations (phonons) or by
the presence of different types of defects randomly distributed across the lattice
\cite{hofmann:chemrev:1996}, for instance, adsorbed particles (adsorbates).
In the case of a periodic surface, the presence of an adsorbate on top of it translates into a blurring of
the well-defined Bragg features and the appearance of broad intensity features.
This diffuse scattering effect \cite{lahee:JCP:1987} is analogous to the image distortion produced by a
rough mirror---the larger the number of CO adsorbates on the clean Pt surface, the larger the distortion
with respect to neat Bragg-like diffraction intensity peaks.

Instead of considering a large number of randomly distributed CO adsorbates, we are going to focus on
the effects produced by a single isolated CO adsorbate.
Moreover, we shall not focus on the quantitative description of the diffraction process itself, but on the
analysis of how the features associated with a given descriptive level manifest on the next level, which
is assumed to be more refined and, therefore, accurate.
Accordingly, we will see that although such features are transferred from the model characterizing one
level to the model corresponding to the next level, it not always easy to establish an unambiguous
one-to-one correspondence.
In simple terms, appealing to a biological metaphor, it is like considering a body and its skeleton.
The skeleton is the structure upon which the body rests.
However, although the body reveals some features of the skeleton (we can perceive the position of some
bones under our flesh), it is a much more complex super-structure.
In particular, here the problem considered is approached at three descriptive levels:
\begin{itemize}
 \item The {\bf Fermatian level}, which refers to the analysis of the problem assuming a bare hard-wall-like (fully repulsive)
  model to describe the He-CO/Pt(111) interaction.
  Because the trajectories here are of the type of sudden impact (free propagation except at the impact point
  on the substrate wall, where the trajectory is deflected according to the usual law of reflection), they are
  going to be straight-like rays, as in optics (this is why it is referred to as Fermatian).

 \item The {\bf Newtonian level}, where the He-CO/Pt(111) interaction is modeled in terms of a potential
  energy surface that smoothly changes from point to point.
  This model has a repulsive wall that avoids He atoms to approach the substrate beyond a certain distance (for a
  given incidence energy), and an attractive tail that accounts for van der Waals long-range attraction.
  The existence of these two regions, repulsive and attractive, gives rise to an attractive channel around the
  CO adsorbate and that continuous along the flat Pt surface, inducing the possibility of temporary trapping
  for the He atoms.

 \item The {\bf Bohmian level} is the upper one and, to some extent, makes an important difference with the
  previous models because here the trajectories are not only dependent on the interaction potential model, but also on the particular
  shape displayed by the wave function at each point of the configuration space at a given time
  (the ``guiding'' or ``pilot''~wave).
\end{itemize}

From a physical view point, we are going to focus on three different aspects or phenomena, namely
{\it reflection symmetry interference}, {\it surface rainbows} and {\i surface trapping}.
As it will be seen, all these aspects have a manifestation in the corresponding diffraction intensity patterns.
{\it Reflection symmetry interference} is the mechanism considered to explain the oscillations displayed by the
intensity pattern at large diffraction angles \cite{lahee:JCP:1987,graham:JCP:1996}, and is based on the
hypothesis that the diffracted wave can be assumed to be constituted by two interfering waves.
In terms of trajectories, as~we shall see, this means that there are pairs of homologous paths (either Fermatian
or Newtonian), which, when using semiclassical arguments in terms of the optical concept of paths difference,
explain the appearance of such type of interference \cite{lahee:JCP:1987,graham:JCP:1996}.
{\it Rainbow} features arise as a consequence of the local changes in the curvature of the interaction potential
model \cite{yinnon:JCP:1988,kleyn:physrep:1991}, which give rise to accumulations of classical trajectories
along some privileged directions (rainbow deflection angles), but~that quantum-mechanically leave some
uncertainties when we look at the corresponding diffraction patterns \cite{lemoine:PRL:1998,sanz:jcp:2004}.
Finally, {\it surface trapping} along the clean Pt surface arises for some deflections from the adsorbate at grazing angles.
This phenomenon takes place when, by virtue of the interaction with the adsorbate, the He atoms lose too much energy
along their perpendicular direction, transferring it to their parallel degree of freedom (perpendicular and parallel are defined
with respect to the clean Pt surface).
The energy associated with their perpendicular degree of freedom becomes negative, while the parallel energy gets larger
than the incident one (by conservation), which ends up with the atom moving in the form of jumps along the surface until it encounters
another adsorbate that, by~means of the reverse process, can be used to release the atom from the surface \cite{glebov:PRL:1997}.

According to the above discussion, this work has been organized as follows.
Details about the interaction potential models considered to determine the different trajectory dynamics
are provided and discussed in Section~\ref{sec2}.
In this section, a brief discussion is also given concerning the numerical approaches used to determine
the calculations shown here.
The~diffraction intensity patterns for both the hard-wall model and the realistic interaction
potential are shown and discussed in Section~\ref{sec3}.
Section~\ref{sec4} is devoted to the description and analysis of the different types of trajectories (Fermatian, Newtonian
and Bohmian), emphasizing the features that are both common to all approaches and also their main differences.
To conclude, a series of final remarks are summarized in Section~\ref{sec5}.


\section{Potential Model and Computational Details}
\label{sec2}

Interaction potential models for systems like the one we are dealing with here are determined from
information extracted from experimental diffraction patterns.
Thus, let us consider that, as it is typically done, the intensity distribution in these patterns
is specified in terms of the transfer of He-atom momentum from the incidence direction to the
direction parallel to the surface or, in brief, parallel momentum transfer \cite{lahee:JCP:1987}, i.e.,
\be
 \Delta K = k_{d,x} - k_{i,x} = k_i (\sin \theta_d - \sin \theta_i) ,
 \label{pmt}
\ee
where $\theta_d$ and $\theta_i$ are the diffraction and the incident angles, respectively, and
$k_i = \sqrt{2mE_i/\hbar}$ is the incident wavenumber.
Then, diffraction features in the large-angle region of the pattern (large parallel momentum transfers) typically convey
information about the repulsive part of the interaction, while the attractive part has a more prominent influence on the small-angle
region (low parallel momentum transfers).
From these potential models, it is possible to estimate the effective size of the adsorbed particles \cite{lahee:PRL:1986,lahee:JCP:1987}
as well as the local curvature of the surface electron density, which additionally may induce the presence and contribution of rainbow-like
features \cite{child-bk:1974,kleyn:physrep:1991} whenever the interaction potential consists of a short range repulsive region followed
by a longer range attractive one (accounting for the van der Waals interaction between the adsorbate-surface system and the impinging
neutral atom).

\begin{figure}[H]
 \centering
 \includegraphics[width=9cm]{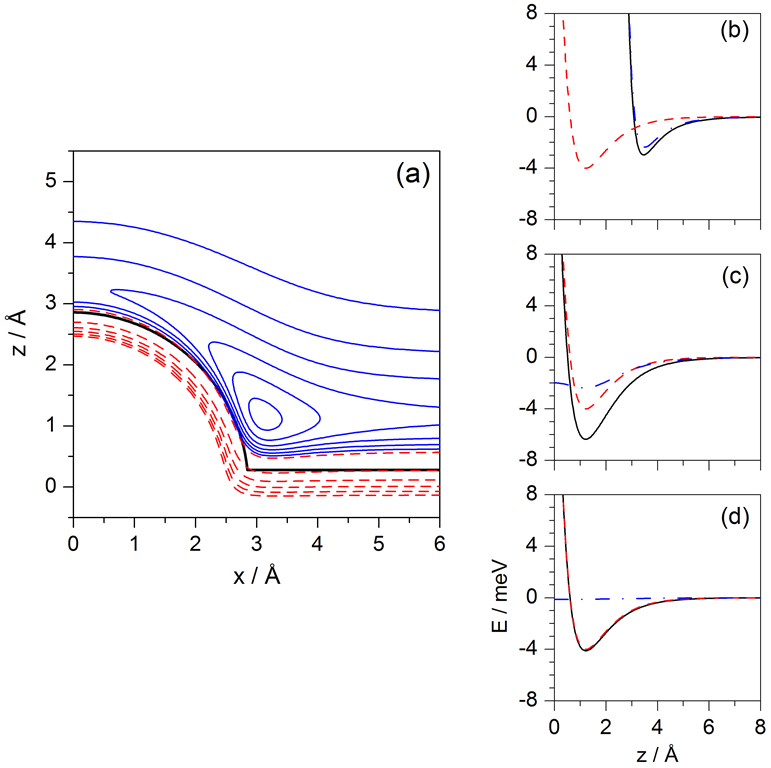}
 \vspace{-6pt}
 \caption{(\textbf{a}) contour plot of the He-CO/Pt(111) interaction potential model (\ref{eq:1.1}) (see text for details).
  The~energy difference between consecutive repulsive/attractive contour levels (red dashed lines/blue solid lines) is 10~meV/1~meV.
  The thick black solid line denotes the repulsive boundary for an approximate hard-wall model set for an incidence energy of 10~meV
  (see text for details).
  In the right-hand side panels, energy profiles along the $z$ direction for:  (\textbf{b}) $x = 0$~\AA; (\textbf{c}) $x = 3.31$~\AA\ and (\textbf{d}) $x = 6.35$~\AA.
  In~these panels, the total interaction potential is denoted with black solid line, while red dashed and blue dash-dotted lines refer to the
  Morse and Lennard--Jones contributions, respectively.}
 \label{Fig1}
\end{figure}

The model considered here, proposed by Yinnon {et al.}\ \cite{yinnon:JCP:1988}, gathers the above mentioned features
and nicely fits the experimental data \cite{yinnon:JCP:1988,lemoine:JCP:1994-1,lemoine:PRL:1998}.
This potential model consists of two terms:
\begin{equation}
 V({\bf r}) = V_{\rm Pt(111)}(z) + V_{\rm CO}({\bf r}) ,
 \label{eq:1.1}
\end{equation}
where $V_{\rm Pt(111)}(z)$ and $V_{\rm CO}({\bf r})$ describe, respectively, the
separate interaction of the He atom with the clean Pt surface and the adsorbate.
The functional form of these two potential functions are a Morse potential for the
He-Pt(111),
\begin{equation}
 V_{\rm Pt(111)} (z) = D
  \left[ 1 - {\rm e}^{- \alpha (z - z_m )} \right]^2 - D ,
 \label{eq:1.2}
\end{equation}
and a Lennard--Jones potential for the He-CO,
\begin{equation}
 V_{\rm CO} ({\bf r}) = 4 \epsilon \left[
  \left( \frac{\sigma}{r} \right)^{12}
   - \left( \frac{\sigma}{r} \right)^6 \right] ,
 \label{eq:1.3}
\end{equation}
with $r = \sqrt{x^2 + y^2 + z^2}$.
The reference system for the potential (\ref{eq:1.1}) is centered at the CO center of mass, with
$D = 4.0$~meV, $\alpha = 1.13$~\AA$^{-1}$, $z_m = 1.22$~\AA\ and $\epsilon = 2.37$~meV
\cite{lemoine:JCP:1994-1,lemoine:PRL:1998}.
With this, $z$ accounts for the position of the He atom along the perpendicular direction with respect
to the clean Pt surface, while $x$ and $y$ describe its parallel motion.
Because of the rotation symmetry around the axis $x=y=0$ exhibited by the interaction potential
(\ref{eq:1.1}), for an illustration, in the contour plot displayed in Figure~\ref{Fig1}a only one half of the
transverse section along the plane $y=0$ is shown.
In it, negative and positive energy contours are denoted, respectively, with blue solid line and red dashed
line (due to symmetry, only a half of the potential is shown).
On the right-hand side, panels (b)--(d) show different transverse sections of the potential to better
appreciate the effect of the local curvature along the $z$ direction at three different distances from $x=0$
(for $x=0$~\AA, $x=3.31$~\AA\ and $x=6.35$~\AA, respectively), which give an idea of the well-depth on top of
the adsorbate (about 2.96~meV), at the intersection of the adsorbate with the flat Pt surface (6.37~meV),
and on top of the flat surface far from adsorbate (4~meV),~respectively.

The existence of an attractive well around the adsorbate and also along the surface is going to induce
temporary trapping both classically and quantum-mechanically---only the presence of another
neighboring adsorbate may help to remove such trapping.
Since far from the influence region of the adsorbate the He-Pt(111) interaction only depends on the
$z$ coordinate, the trapped motion will be ruled by the well of the Morse function, being free along
the $x$ direction.
The resulting motion is thus a combination of jumps along the $z$ direction with a uniform motion parallel
to the Pt surface, with an average speed larger than $\sqrt{2E_i/m}$, with $E_i$ being the incident energy
(notice that the energy along the $x$ direction has to be larger than along the $z$ direction in order to
achieve negative values along the latter and, hence, trapping).
Classically, if the energy along the $z$ direction is given by
\be
 E_z = \frac{p_z^2}{2m} + V_{Pt(111)} (z) ,
 \label{eq:1.4}
\ee
where $E_z = E_i - E_x < 0$ and $m$ is the He-atom mass, it is easy to show that the turning points will be located at
\be
 z_\pm = z_m - \frac{1}{\alpha} \, \ln \left[ 1 \pm
         \sqrt{1 - \frac{|E_z|}{D}} \right] .
 \label{eq:1.7}
\ee
This motion is anharmonic, with frequency
\be
 \omega = \sqrt{\frac{2 \alpha^2 |E_z|}{m}} .
 \label{eq:1.9}
\ee
The length of each jump along the $x$ direction can be easily estimated from {Equation} (\ref{eq:1.9}) according to the~relation
\be
 \Delta x = \frac{p_x}{m} \ \tau
  = \frac{2\pi}{\alpha}\ \sqrt{\frac{E_x}{(E_x - E_i)}}
  = \frac{2\pi}{\alpha}\ \sqrt{\frac{(E_i - E_z)}{|E_z|}} ,
 \label{eq:1.10}
\ee
where $p_x = m v_x$ and $\tau = 2\pi/\omega$.
As can be noticed, at the threshold $E_x = E_i$, the jump length becomes infinity, i.e., the trajectories
leave the adsorbate being and remaining parallel to the clean Pt surface.

Quantum-mechanically, the trapped portions of the wave packet will somehow contain information about
the bound states of the Morse function, with eigenenergies given by \cite{morse:physrev:1929}
\be
 E_n = \hbar \Omega \, \left( n + \frac{1}{2} \right) \,
  \left[ 1 - \frac{\hbar \Omega}{4D} \, \left( n + \frac{1}{2} \right) \right] ,
 \label{eq:1.11}
\ee
with
\be
 \Omega = \sqrt{\frac{2 \alpha^2 D}{m}}
 \label{eq:1.12}
\ee
being the harmonic frequency resulting from approximating the Morse potential to
a harmonic oscillator.
From the condition to determine the total number of bound states, i.e.,
\mbox{$\Delta E_{n'=n+1,n} = E_{n'} - E_n \ge 0$}, it is found that, for the parameters
here considered, there are only three bound states, namely the ground state plus
two excited ones: $E_0 = - 2.53$~meV, $E_1 = - 0.60$~meV and $E_2 = - 3 \times 10^{-3}$~meV.
If~we assume $E_z = E_n$, then we obtain $\Delta_x^{(0)} \approx 12$~\AA, $\Delta_x^{(1)} \approx 23$~\AA\
and $\Delta_x^{(2)} \approx 320$~\AA, respectively, according to Equation~(\ref{eq:1.10}).

The potential model (\ref{eq:1.1}) will be used to investigate the behavior of Newtonian and Bohmian
trajectories when the He atoms are influenced by it.
For the Fermatian approach, we shall consider a rather crude approximation to this potential, which consists of
substituting it by a purely repulsive infinite barrier, which will be referred to as the (repulsive) hard-wall model.
This barrier is an approximation to the equipotential $V({\bf r}) = 10$~meV (black thick solid line in Figure~\ref{Fig1}a),
so that $V_{HW}({\bf r}) = \infty$ for any position ${\bf r}$ such that $V({\bf r}) \ge 10$~meV and zero everywhere else
(i.e, for all ${\bf r}$ such that $V({\bf r}) < 10$~meV).
An optimal fit to this equipotential renders a radius $a = 2.86$~\AA\ for the adsorbate and a position $z_r = 0.28$~\AA\
of the clean Pt surface above the CO center of mass.
Since the Bohmian treatment of this model is not included here, it is worth mentioning a recent analysis of an analogous
system by Dubertrand {et al.}~\cite{struyve:JPA:2018}, where the authors consider the Bohmian description of quantum
diffraction by a half-flat surface (simplified by a half-line barrier), earlier considered by Prosser \cite{prosser:ijtp:1976-1}
in the context of the solution of Maxwell's equations to the problem of diffraction by a semi-infinite conducting sheet
\cite{bornwolf-bk}.

One of the advantages of the model introduced in this section is that it can describe the presence of a
localized adsorbed particle on a surface (which can be regarded as a point-like defect \cite{lahee:JCP:1987}),
but~also a row of adsorbates aligned, for instance, along the $y$-axis (a linear-like defect \cite{lahee:PRL:1986}).
In the first case, we have radial symmetry with respect to the $z$-axis, while the latter is characterized by axial
symmetry, along the $y$-axis.
From the viewpoint of a trajectory-based description, there is no difference between one case and the other, since
what happens in one half of a transverse section also happens in any other section (which can be reconstructed
either by rotation symmetry around the $z$-axis or by translational symmetry along the $y$-axis and/or mirror
symmetry with respect to the $yz$ plane).
The difference between both models relies on the way how the trajectories distribute spatially and therefore how many
of them lay within a certain solid angle, independently of whether the trajectories are classical (Fermatian or
Newtonian) or quantum-mechanical (Bohmian).
Having said this, since we are interested in comparing the behavior of different types of trajectories, from now on,
the discussion will turn around the two-dimensional description, which is analogous to consider an axial-symmetric~system.

Regarding the conventions used here \cite{sanz:SSR:2004}, the incidence angle, $\theta_i$, for Fermatian
and Newtonian trajectories is defined as the angle subtended between the incident direction of a given
trajectory and the normal to the clean Pt surface.
Dynamically speaking, this translates into an effective way of how much momentum is provided initially to
each direction, i.e.,
\be
\begin{array}{rcl}
 p_{i,x} & = & p_i \sin \theta_i = \sqrt{2mE_i} \sin \theta_i , \\
 p_{i,z} & = & - p_i \cos \theta_i = - \sqrt{2mE_i} \cos \theta_i ,
\end{array}
\label{momenta}
\ee
%
%
with $p_i = \hbar k_i = \sqrt{2mE_i}$.
In particular, in the calculations presented and discussed below, we have considered two values of the incident
energy, namely $E_i = 10$~meV and 40~meV.
In terms of the de Broglie wavelength, $\lambda_{\rm dB} = h/\sqrt{2mE_i}$, with $h$ being Planck's constant,
these energies correspond to $\lambda_{\rm dB} = 1.43$~\AA\ and 0.72~\AA, respectively.
The relations (\ref{momenta}) are also used for the quantum analysis, where the incident wave function is launched
with a momentum in compliance with these expressions.
The deflection (or outgoing) angle, $\theta_d$, on the other hand, is defined as the angle subtended
by the normal and the deflection direction for the corresponding trajectory---in the case of wave-function
descriptions, this angle is going to be denoted as the diffraction angle ---, although an analogous definition
in terms of the momentum with which the particle is deflected can also be used.
Once the incidence angle is established, depending on the initial position assigned to the trajectories, they
will behave in a way or another.
To characterize the trajectories according to their initial positions, it is common to refer them to the impact
parameter, which in the present context is defined as the impact position on the clean surface in absence of interaction.
In periodic surfaces, the range for impact parameters ($b$) is typically established in terms of the lattice parameter
(the unit cell length) \cite{sanz:SSR:2004,sanz:PhysRep:2007}.
Here, because the presence of the adsorbate breaks the periodicity of the clean surface, we need to redefine this
range in an alternative and slightly different way.
Specifically, this range is taken as a portion of surface that covers the
extension of the adsorbate and goes well inside the region where the flat surface potential is already not influenced by the adsorbate attractive tail.
With the potential function used here, this~means is satisfied by impact parameters taken within the
range $[-10.6,10.6]$~\AA.
Accordingly, the initial positions are specified as
\be
\begin{array}{rcl}
 x_i & = & b - z_i \tan \theta_i , \\
 z_i & = & z_0 ,
\end{array}
\label{positions}
\ee
%
%
with $b \in [-10.6,10.6]$~\AA, and where $z_0= 10.27$~\AA\ has been chosen far from the surface, such that
$V({\bf r}) \approx 0$ (the same holds for an incident wave function, whose probability density has to be far
from the influence region of the adsorbate).

With the above definitions, the computation of Fermatian trajectories is trivial, since we only need to know
their incidence direction and, from it, the point on the substrate (adsorbate or flat surface) where they will
feel the impact.
At such a point, applying the law of reflection, we readily obtain the deflection angle and therefore the deflected
part of the trajectory.
Fermatian trajectories are just a pure geometric issue and, as it will be seen in next section, the corresponding
quantum calculations are just analytical, so they do not imply high computational demands.
For Newtonian trajectories, however, the computational task is more refined, since the action of the interaction
potential introduces important changes in the curvature of the trajectories when they are approaching the
substrate.
Nevertheless, the computational demand is still low, since such trajectories can be readily obtained by integrating
Newton's equations (actually, Hamilton's equations) with a simple fourth-order Runge--Kutta algorithm using the
above momentum and position values, \mbox{Equations~(\ref{momenta}) and (\ref{positions})}, respectively, as initial conditions.

The numerical computation of the wave-function evolution and the Bohmian trajectories is, however, more subtle,
since it implies the solution of a partial differential equation.
In this case, integration has been carried out by means of the second-order finite-difference algorithm
\cite{leforestier:jcompphys:1991}, making use of the fast Fourier transform to compute the kinetic part
of the operator \cite{press-bk-1}.
For the initial wave function, we would like to simulate an incident nearly plane wave, which mimics a highly
collimated He-atom beam.
Numerically, we can recreate this situation by considering a quasi-monochromatic wave function or wave packet that
covers the substrate well beyond the effective size of the adsorbate.
This can be done by linearly superimposing a large number of Gaussian wave packets, which in our cases amounts
to considering 250 Gaussian wave packets \cite{sanz:JPCM:2002}, where the spreading of each wave packet is
0.84~\AA\ along the $x$ direction and 2.65~\AA\ along the $z$ direction.
With these conditions, the wave function reaches the surface with almost no increase of its size, which is launched
from an average position along the $z$ direction $\langle z \rangle_0 = z_0 = 10.27$~\AA\ and normal incidence
conditions (again, for visual clarity).
In order to ensure an optimal overlapping along the $x$ direction, the centers of the wave packets
are separated a distance of 0.21~\AA.
For simplicity both here and also with the hard-wall model, the quantum calculations have been performed
at normal incidence conditions ($\theta_i = 0^\circ$), although this does not diminish the generality of
the results presented.

Regarding the computation of the Bohmian trajectories, they are synthesized on the fly from the wave function.
That is, the wave function is made to evolve for a small time interval $dt$, and then the trajectories are propagated
from their actual position to the new one with the phase information provided by the updated wave function.
The equation of motion that rules this behavior is the guidance Equation (\ref{eqbohm}) (see Section~\ref{sec43}
for further details on this equation of motion).
Since the value of the wave function and its derivatives is known only on the knots of the numerical grid, the guidance
equation has to be solved with the aid of numerical interpolators, which render the values required at any other point
other than a grid knot with a reliable accuracy.
With these values, the equation of motion is solved by means of a Runge--Kutta algorithm, as in the case of Newtonian
trajectories, although the degree of accuracy required is higher, particularly due to the appearance of nodal structures.
As for the initial conditions, they have been chosen along lines parallel to the flat surface, at different constant distances
from the latter and taking the value $\langle z \rangle_0 = z_0$ as a reference.
Specifically, three of these lines have been taken above this value [$z(0) > z_0$] and another three have been
taken below [$z(0) < z_0$], i.e., closer to the substrate (see Section~\ref{sec43} for further details).


\section{Wave-Function Approach}
\label{sec3}
\vspace{-6pt}


\subsection{Diffraction from a Repulsive Hard-Wall Potential}
 \label{sec31}

In the case of the two-dimensional (axial-symmetric) version of the hard-wall model described
in Section~\ref{sec2}, the diffracted wave far from the adsorbate can be obtained from the exact
(analytical) asymptotic solution to the problem of the diffraction from a cylinder \cite{morse-bk-1},
\be
 \Psi = e^{i {\bf k}_i \cdot {\bf r}} + \frac{f ({\bf k}_d)}{\sqrt{r}} \ e^{i k_d r}
  = e^{i {\bf k}_{i,x} x - k_{i,z} z} + \frac{f (k_{d,x},k_{d,z})}{\sqrt{r}} \ e^{i k_d r} ,
 \label{eq:2.3}
\ee
with ${\bf r} \to \infty$, and where ${\bf k}_i = (k_{i,x}, k_{i,z})$ is the incidence wave vector
(momentum) and ${\bf k}_d = (k_{d,x}, k_{d,z})$ is an outgoing wave vector pointing along an arbitrary diffraction
direction---the two wave vectors are expressed in terms of their parallel and perpendicular components
($k_{i,x}$ and $k_{d,x}$, and $k_{i,z}$ and $k_{d,z}$, respectively) on purpose.
In this expression, the first term is the contribution from the direct wave and the second term
accounts for the diffraction caused by the defect itself (the minus sign in the first contribution
arises from the fact that the incidence direction is considered to be negative).

If instead of a cylinder we have a half of it on top of a flat surface, the solution (\ref{eq:2.3}) has to include the
effect of the reflection from the flat surface, i.e.,
\be
 \Psi = e^{i (k_{i,x} x - k_{i,z} z)} - e^{i (k_{i,x} x + k_{i,z} z)} + \frac{f (k_{d,x},k_{d,z})}{\sqrt{r}} \ e^{i k_d r} ,
 \label{eq:2.4}
\ee
which has to satisfy the hard-wall boundary conditions
\bd
 \Psi (x,z = z_r) = \Psi (r = a) = 0 .
\ed
The first two terms in (\ref{eq:2.4}) satisfy this condition when $z = 0$, as expected on the flat surface.
On the other hand, for the third term to satisfy these boundary conditions it is necessary that the diffraction
(or scattering) amplitude, $f$, is given by two contributions,
\be
 f (\theta_i, \theta_d) = f_a (|\theta_d - \theta_i|) - f_b (\pi - |\theta_d + \theta_i|) ,
 \label{eq:2.5}
\ee
where the first term describes direct reflection from the adsorbate and the
second, a double reflection from the adsorbate and then the flat surface.
Both amplitudes can be recast in terms of the difference between the diffraction
and incidence angles.
In the first case, this can be readily seen; in the latter, a~similar result is obtained after assuming
collisions with a full cylinder, because then the diffraction angle is $\theta'_d = \pi - \theta_d$.
In addition, note that the symmetry displayed by {Equation} (\ref{eq:2.5}) for the specific case
$\theta_i = 0$ is analogous to the antisymmetry condition arising in fermion-fermion collisions
\cite{joachain-bk}, where the symmetrized amplitude for two fermions with $1/2$-spin
in a triplet state has the functional~form
\be
 f_{-} (\theta) = \frac{1}{\sqrt{2}} \left[ f (\theta) - f (\pi - \theta) \right] .
 \label{eq:2.6}
\ee

Analytically, in the short-wavelength limit ($ka \to \infty$) and for a cylindrical defect,
the diffraction amplitudes in {Equation} (\ref{eq:2.5}) are of the form \cite{morse-bk-1},
%
\be
 f (\theta) =
  - \left[ \frac{a \, {\rm sen} \, \theta/2}{2} \right]^{1/2} \,
        {\rm e}^{- 2 {\rm i} k a \, {\rm sen} \, \theta/2}
  + \frac{{\rm e}^{-{\rm i}\pi/4}}{\sqrt{2\pi k}}
    \frac{( 1 + \cos \theta )}{{\rm sen} \, \theta} \,
    {\rm sen} \, (ka \, {\rm sen} \, \theta ) .
 \label{eq:2.8}
\ee
where the first term, known as the illuminated face term, accounts for the backward
scattering of the wave and the second one describes the Fraunhofer diffraction.
The final analytical expression for the diffraction amplitude $f$ is obtained by
considering the symmetrization condition (\ref{eq:2.5}) in these results.

\begin{figure}[H]
 \centering
 \includegraphics[width=10cm]{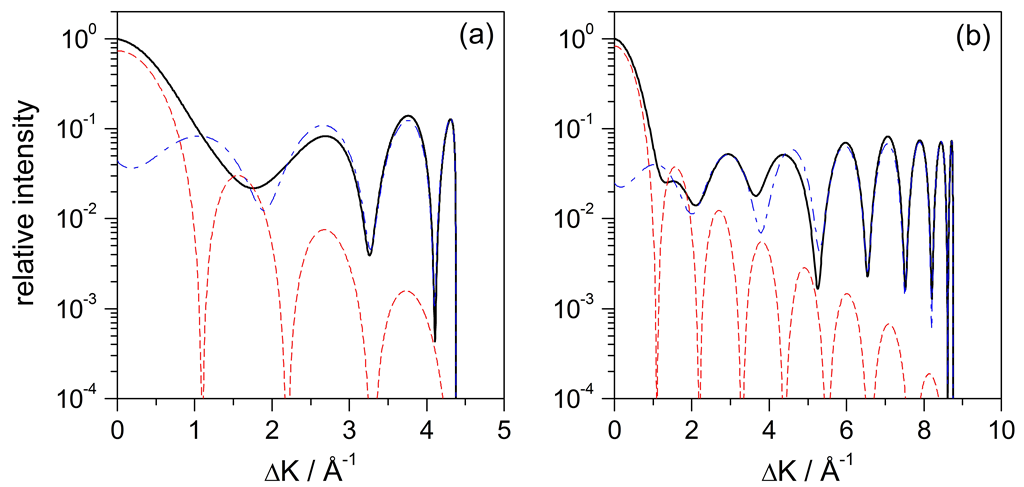}
 \caption{(\textbf{a}) Relative diffraction intensity (black solid line) produced by a radial hard-wall model for incidence
  conditions $\theta_i = 0^\circ$ and $E_i = 10$~meV.
  To compare with, the Fraunhofer and illuminated-face intensities are also shown, which are denoted with
  red dashed line and blue dash-dotted line, respectively;
  (\textbf{b}) As in panel (\textbf{a}), but for $E_i = 40$~meV.}
 \label{Fig2}
\end{figure}

The diffraction intensity for the axial-symmetric, fully repulsive hard-wall model is shown in
Figure~\ref{Fig2}a as a function of the parallel momentum transfer (\ref{pmt}), for $E_i = 10$~meV and
normal incidence ($\theta_i = 0^\circ$).
In the same figure, the intensities associated with the illuminated face term and the Fraunhofer diffraction
are also displayed separately (red dotted line and blue dashed line, respectively) in order to get a better
idea at a quantitative level of their respective contributions to the total pattern.
As it can be seen, for small momentum transfers (small diffraction angles), the leading term is the
Fraunhofer one, which decreases fast as the momentum transfer increases (as $(\Delta K)^{-2}$).
On the contrary, the illuminated-face term, together with its mirror image, becomes the leading contribution
for large momentum transfers (large diffraction angles).
The type of oscillations generated by these two terms, the illuminated-face one and its mirror image, give
rise to the reflection symmetry phenomenon \cite{lahee:JCP:1987}, which explains why the diffraction pattern
does not decay for large momentum transfers, as happens, for instance, in simpler cases of diffraction by a
wire or a slit.
This behavior is observed regardless of the incident energy, as can be noticed from the intensities displayed
in Figure~\ref{Fig2}b for $E_i = 40$~meV (and~also normal incidence).

\subsection{Diffraction from the Potential Model (\ref{eq:1.1})}
\label{sec32}

The quantum treatment for the potential model (\ref{eq:1.1}) does not admit analytical solutions, as it
is the case of the hard-wall model of Section~\ref{sec31}, which provides us with an asymptotic analytical
solution of the diffraction far from the interaction region.
A way to tackle the problem is by using a numerical wave-packet propagation method, as described in
Section~\ref{sec2}, which renders a description of the diffraction phenomenon in real time, i.e., providing
us with direct information on the time-evolution of the He-atom wave function, of particular interest in
the region where the interaction between the He atom and the substrate is stronger.
Hence, although we lose analyticity, we gain insight on the dynamical process in the interaction region.

\begin{figure}[H]
 \centering
 \includegraphics[width=11cm]{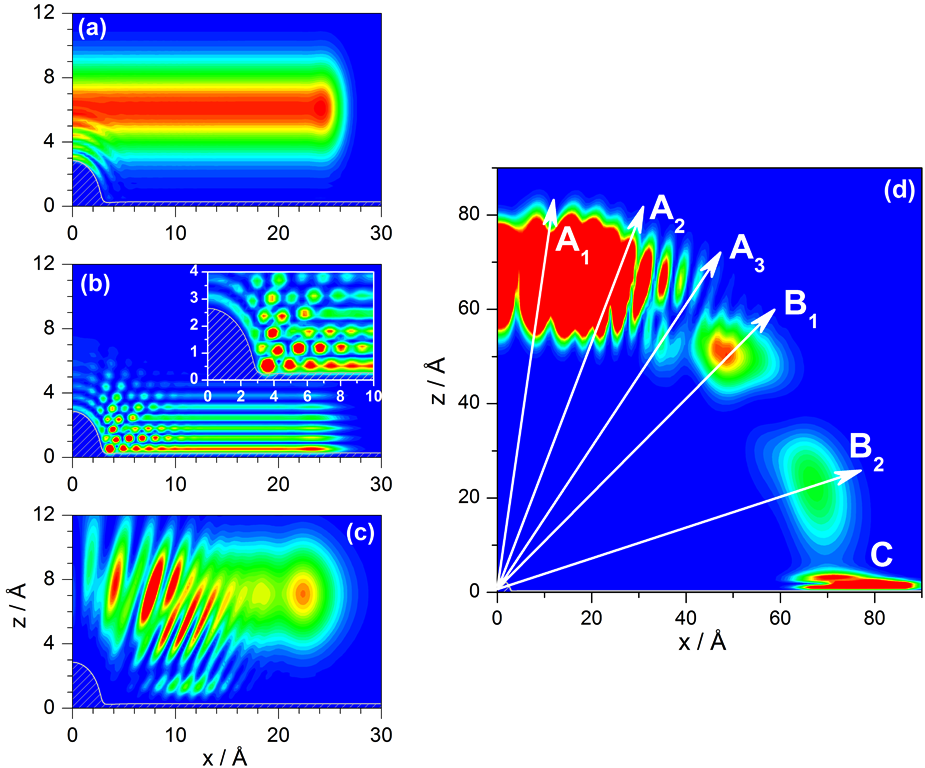}
 \caption{On the left-hand side, contour plots illustrating three different instants of the evolution of the
  probability density near the surface for incidence conditions $\theta_i = 0^\circ$ and $E_i = 10$~meV:
  (\textbf{a}) when the density starts being influenced by the adsorbate; (\textbf{b}) when the density is totally interacting
  with the substrate (i.e., with both the adsorbate and the flat Pt surface) and (\textbf{c}) when the density starts
  leaving the substrate.
  In panel (\textbf{d}), on the right-hand side, plot of the probability density far from the influence of the adsorbate ($t=11$~ps).
  Arrows and capital letters denote different diffraction directions to be identified in the intensity plot displayed in
  Figure~\ref{Fig4}a: $A_i$: directions identifying interference features associated with the superposition of
  the circular and planar wavefronts, contributing to the central maxima of the intensity pattern;
  $B_i$: associated with features arising from the reflection symmetry interference phenomenon;
  $C$: surface trapping.}
 \label{Fig3}
\end{figure}

Accordingly, the evolution of the probability density as it approaches the adsorbate and then gets diffracted
is shown in Figure~\ref{Fig3} at three different instants of its evolution for $E_i = 10$~meV and normal
incidence (taking advantage of the mirror symmetry with respect to $x=0$ due to the normal incidence, only
a half is plotted for simplicity).
In panel (a), we observe the appearance of circular wavefronts (ripples) around the adsorbate due to the
interference produced by the overlapping of the part of the wave function that is still approaching the
adsorbate with the part that is already being diffracted.
In panel (b), the whole of the wave function is interacting with the substrate (at about 1.5~ps).
In this case, there are circular wavefronts around the adsorbate, as before, but also additional plane wavefronts
produced by an analogous interference process associated with the portion of the flat Pt surface reached by the
incident wave function.
The superposition of the circular wavefronts with the planar ones generates around the adsorbate a web of maxima
(see inset for a more detailed picture), which gets weaker and blurred as we move far from the adsorbate.
The periodicity of the web of maxima is related to the incoming He-atom de Broglie wavelength,
$\lambda_{dB} = 1.43$~\AA, although with some distortions due to the presence of the attractive
region of the interaction.
As the wave function further evolves, we can observe a superposition of two contributions, as seen in panel (c).
One of them is the reflection of a nearly square function, which starts displaying the typical Fresnel or near-field
features of such functions when they arise from single slit diffraction.
This contribution is precisely the illuminated face term that we saw in Section~\ref{sec31}.
The other contribution, which distributes around the adsorbate, is going to be related to the Fraunhofer diffraction
and also to other features, such as trapping of the wave along the attractive well near the Pt surface.
Nonetheless, notice that the final diffraction pattern, as in the case of the hard-wall model (see Equation~(\ref{eq:2.8})),
is a superposition of the two contributions (diffraction amplitudes), and not only the direct sum of their probabilities.

\vspace{-3pt}

\begin{figure}[H]
 \centering
 \includegraphics[width=9cm]{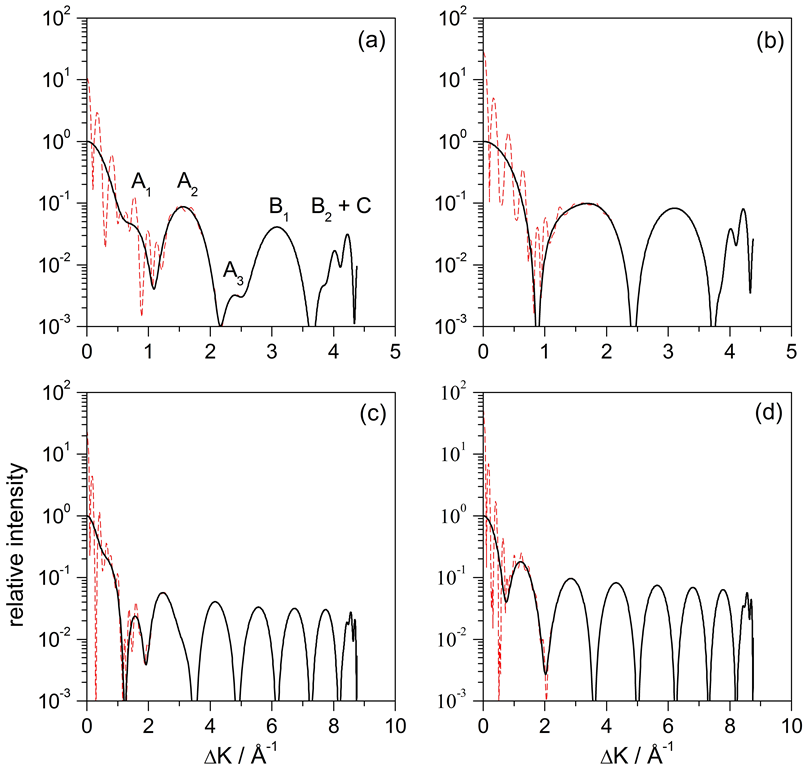}
 \caption{(\textbf{a}) Relative diffraction intensity (black solid line) produced by an axial-symmetric potential model
  based on (\ref{eq:1.1}) for incidence conditions $\theta_i = 0^\circ$ and $E_i = 10$~meV.
  For comparison, the intensities before (red dashed line) and after (black solid line) removing the plane-wave
  contribution (see text for details) are both plotted;
  (\textbf{b}) the same as in panel (\textbf{a}), but for a fully repulsive model of the CO adsorbate, obtained after removal of the
  attractive part of the Lennard--Jones function (\ref{eq:1.3}).
  To compare, in panels (\textbf{c},\textbf{d}), the same as in panels (\textbf{a},\textbf{b}), respectively, but for $E_i = 40$~meV.}
 \label{Fig4}
\end{figure}

Asymptotically, far from the influence of the classical interaction, as shown in Figure~\ref{Fig3}d,
it is possible to observe traits related to the evolution of the three parts of the scattered wave, which
can be eventually associated with the peaks characterizing the corresponding intensity pattern (see
Figure~\ref{Fig4}a).
Thus, interferences arising from the superposition of the circular and planar wavefronts give rise to
the central features of the intensity pattern (denoted with directions labeled with $A_i$ in the plot).
However, there are also other types of peaks, which propagate along the directions denoted as $B_i$ and,
as will be seen below, can be associated with the reflection symmetry interference phenomenon.
Both~types of peaks, $A_i$ and $B_i$, implicitly carry information about the classical rainbow (see
Section~\ref{sec42}) in a sort of global fashion, since this phenomenon does not manifest with a particular
well-defined peak, in~general.
On the other hand, it is also possible to observe the presence of trapping ($C$), which is related to
the lower part of the wave function that keeps moving close and parallel to the clean Pt~surface.

\vspace{-3pt}
In order to quantify the effects produced by the diffraction process in the incoming wave, we represent the
diffracted wave function as a linear combination of plane waves,
\be
 \Psi ({\bf r},t) \sim \int \frac{S(k_{i,x},k_{d,x})}{\sqrt{k_{d,z}}}\ e^{i(k_{d,z} z + k_{d,x} x)} dk_{d,x},
\ee
where the elements $S(k_{i,x},k_{d,x})$ provide the probability amplitudes associated with the change
of parallel momentum $\Delta K = k_{d,x} - k_{i,x}$.
These elements are determined by projecting the numerically computed diffracted wave onto this expression,
\be
 S(k_{i,x},k_{d,x}) \sim \int \sqrt{k_{d,z}} e^{-i {\bf k}_d \cdot {\bf r}} \Psi({\bf r},t) d{\bf r},
\ee
from which the probability $|S(k_{i,x},k_{d,x})|^2$ is obtained to detect atoms that have exchanged
a given amount $\Delta K$ of parallel momentum is obtained (i.e., the reflection coefficient).
With periodic lattices, this calculation is typically performed within a single unit cell; here, because of the lack of periodicity,
the calculation involves an artificial cell of about 53~\AA, which covers a region large enough as to include both the adsorbate and
a good portion of the flat surface that is not influenced by artifacts related to the adsorbate curvature---this is appropriate
to capture isolated signatures of trapping (otherwise, there could be some contamination from the wave scattered in other
directions).
This procedure, however, has an inconvenience: the intensity pattern includes a rather high contribution from mirror reflection
from the flat Pt surface.
In order to remove it, the incident plane wave, ranging from $-\infty$ to $\infty$, is~decomposed as a linear superposition of
two contributions, one contained within the integration range $\mathcal{R}$ ($\phi$) and another outside it ($\chi$), i.e.,
\be
 \Psi_0 ({\bf r}) = \sum_{x_i \in \mathcal{R}} \phi^i ({\bf r}) + \sum_{x_i \notin \mathcal{R}} \chi^i ({\bf r})
  = \sum_{x_i \in \mathcal{R}} \left[ \phi^i ({\bf r}) - \chi^i ({\bf r}) \right]
  + \sum_{\forall \, x_i} \chi^i ({\bf r}),
 \label{eq:4.1}
\ee
where the subscript $i$ labels functions from a given basis set established on purpose to construct
the wave function.
Because the $\chi$ waves essentially describe diffraction from the clean Pt surface, the last
term in this expression is going to be a contribution with the form of a $\delta$-function along
the incident direction.
Thus neglecting this contribution, if the diffraction process from the surface is altered, we make a projection
of the wave on the plane-wave basis set, and the $S$-matrix elements can be recast as
\be
 S(k_i, k_d) = S^\mathcal{R} (k_i, k_d) -  S^\mathcal{R}_0 (k_i, k_d),
 \label{eq:4.2}
\ee
where $S^\mathcal{R}_0 (k_i, k_d)$ is the matrix element corresponding to the scattering of the wave
by a flat unit cell.
The intensity to be compared with the experiment is thus obtained from the differential
cross section or differential reflection coefficient,
\be
 \frac{dR}{d\theta_d} \propto k_{d,z} |S(k_i, k_d)|^2.
 \label{eq:4.4}
\ee
This intensity is displayed in Figure~\ref{Fig4}a as a function of the parallel momentum transferred
(instead of the deflection angle, $\theta_d$) after the wave function has evolved for 5.5~ps after
the maximum approach to the surface.
By this time, the action of the adsorbate interaction potential on the wave function is already
negligible.
Comparing the solid line with the dashed line, we notice the effect of removing the contribution
from the plane wave.
Accordingly, using (\ref{eq:4.2}) is analogous to ``smoothing'' the diffraction pattern, where the
probability distribution does not show well defined maxima because of the presence of the plane-wave
contribution.
The clean oscillations that we observe (solid line) arise from the interference between the circular
wave fronts coming from the adsorbate and the plane wavefronts coming from reflection from the
flat Pt surface.
This ``interaction'' is more prominent for small values of $\Delta K$, this being the reason why the
pattern, before removing the plane-wave contribution, displays fast oscillations.
Such oscillations, however, get weaker and even meaningless as $\Delta K$ increases, because
the circular wavefronts become less affected by their overlapping with the flat outgoing wavefronts.
This is analogous to the behavior already observed with the hard-wall model, in Figure~\ref{Fig2}a:
for small values of $\Delta K$, the dominant contribution was the Fraunhofer one, while, for larger
values of $\Delta K$, the leading one was the illuminated-face contribution.

The intensity maxima in Figure~\ref{Fig4}a, though, do not totally correspond with those
in Figure~\ref{Fig2}a, even if their number is the same.
In particular, notice that some of these maxima ($A_1$ and $A_3$) display a kind of ``wings''.
In order to elucidate their origin, the same calculation has been repeated using a repulsive model for
the adsorbate, which consists of removing the attractive part of the Lennard--Jones function (\ref{eq:1.3}).
The results for this model are displayed in panel (b).
Comparing both models, we find that everything is essentially the same, except precisely for the
presence of such ``wings''.
This result is indeed close to the one displayed in Figure~\ref{Fig2}a for the hard-wall model,
although the maxima are wider, which can be associated with the presence of an attractive well around
the substrate.
In order to determine whether it is an effect or not linked to the incidence energy, the same analysis
was repeated for $E_i = 40$~meV and normal incidence.
From the classical calculations, we conclude that such ``wings'' have to be associated with the presence
of rainbows \cite{kleyn:physrep:1991} (see Section~\ref{sec42}), since this phenomenon is linked to the local
curvature of the interaction potential around the adsorbate.
In this regard, the~attractive well around the adsorbate plays a key role, since its removal makes
rainbow features to disappear even though there is still an attractive region along the clean Pt surface.
Actually, on a more quantitative level, notice that, for instance, for $E_i=10$~meV, the rainbow appears
for $\Delta K \approx 1.89$~\AA$^{-1}$.
In Figure~\ref{Fig4}b, this value of $\Delta K$ is close to the maximum of the second lobe,
which would lead to a distortion of this maximum and the adjacent ones (where the traits
$A_1$ and $A_3$ appear).
In the case of $E_i=40$~meV, there is a classical rainbow at $\Delta K \approx 1.13$~\AA$^{-1}$,
which corresponds to the maximum of the second lobe and, in consequence, this lobe essentially
disappears in the attractive model.
Thus,~the~quantum manifestation of classical rainbow features does not necessarily mean
a contribution in terms of given maximum in the intensity pattern \cite{lemoine:JCP:1994-1,lemoine:PRL:1998},
but it can also be in terms of ``global'' phenomenon that affects the whole pattern
\cite{kleyn:physrep:1991,farias:repprogphys:1998}.
This has actually been a controversial point in the literature when assigning the
origin of the different diffraction maxima \cite{yinnon:JCP:1988,lemoine:JCP:1994-1,lemoine:PRL:1998}.

Finally, regardless of whether the interaction model considers an attractive or repulsive
adsorbate, and also independently of the value of the incident energy, we observe that the
last lobe of the intensity plots in Figure~\ref{Fig4} gathers information of both grazing
deflection and trapping.
In panel (a), for instance, this maximum has been label as $B_2 + C$.
According to Figure~\ref{Fig3}d, the maximum for $B_2$ should appear at
$\Delta K \approx 4.16$~\AA$^{-1}$.
On the other hand, in the same figure, for $C$, we should have a maximum parallel
transfer.
However, the trapped probability is indeed oscillating inside the well while it
moves along the $x$ direction, which makes the corresponding $\Delta K$ value
to fluctuate.
Thus, the probability amplitudes related to $B_2$ and $C$ will display some
interference, which is precisely what we observe in the last lobe of all
calculations presented in Figure~\ref{Fig4}.


\section{Trajectory-Based Description}
\label{sec4}

\vspace{-6pt}
\subsection{Fermatian Level}
\label{sec41}

Although it is a rather crude approximation, the hard-wall model is quite insightful
because it allows for explaining and understanding on simple terms the reflection symmetry interference
phenomenon \cite{lahee:PRL:1986,lahee:JCP:1987,yinnon:JCP:1988,gerber:CPL:1984,choi:JCP:2000,morse-bk-1}
as well as the conditions leading to trapping.
As in geometric optics, the key element is the interpretation of wave phenomena in terms of the phase
difference arising from two different but equivalent paths, where by ``equivalent'' we mean that both
leave the surface with the same deflection (outgoing) angle, even if their journeys close to the surface
are quite different (actually, it is this difference that generates the phase difference).
These geometric rays are what we call here Fermatian trajectories with the purpose to highlight such
optical connotation and, as seen below, in the particular problem we are dealing with here, there are
always homologous pairs of such trajectories.
These pairs are formed by one trajectory that undergoes a single bounce from the substrate before getting
deflected and another trajectory that undergoes two bounces (one with the adsorbate and another with the
flat~surface).

\begin{figure}[H]
 \centering
 \includegraphics[width=7cm]{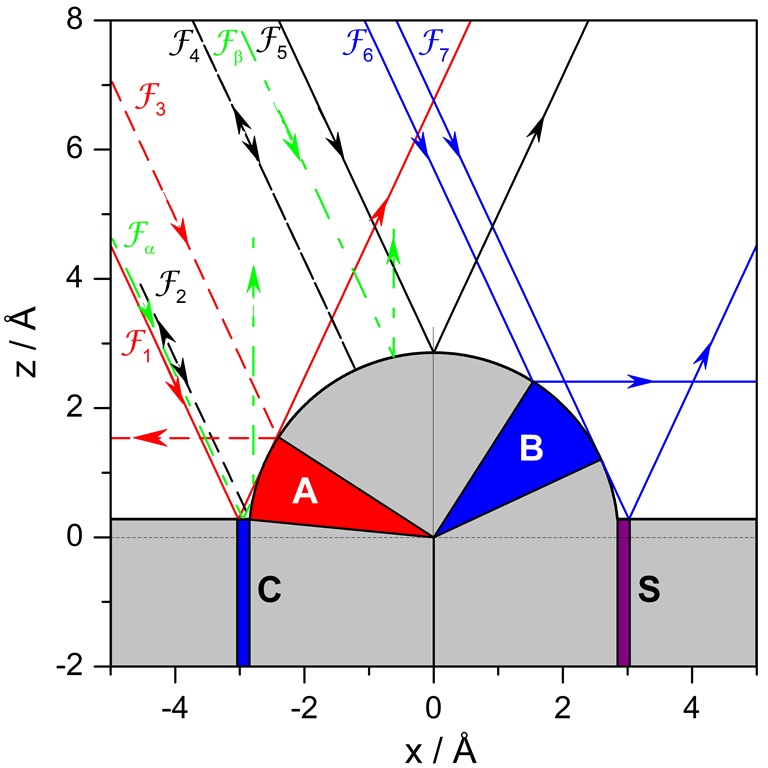}
 \caption{Separatrix Fermatian trajectories for the repulsive hard-wall model set for an incidence energy $E_i = 10$~meV.
  For a better illustration of all possible cases, an incidence angle $\theta_i = 20^\circ$ has been chosen.
  Separatrices displaying forward/backward deflection are denoted with solid/dashed lines.
  Separatrices delimiting regions leading to double scattering in the rear/front part of the adsorbate (regions $C$-$A$/$B$) are
  denoted with red/blue color.
  Separatrices deflected perpendicularly to the flat surface are denoted with green color.
  The shadow region $S$, which depends on the incidence angle, cannot be reached by any trajectory (except for $\theta_i = 0^\circ$,
  where this area goes to zero).}
 \label{Fig5}
\end{figure}

In Figure~\ref{Fig5}, there is a set of Fermatian trajectories, $\mathcal{F}$, of particular
interest: they are separatrices that determine the boundaries for ensembles of Fermatian
trajectories that display a particular behavior, that is, all the Fermatian trajectories
confined within two adjacent separatrices are going to exhibit an analogous behavior.
In general, it can be noticed that, for a given incidence angle (here, $\theta_i~=~20^\circ$),
trajectories may display either a single collision (regions denoted with light gray) or double
collisions (blue and purple regions, denoted with $A$, $B$ and $C$) depending on their impact
parameter.
The~deflection can then be forward (trajectories represented with solid line) or backward
(dashed line).
Moreover, there are also regions of geometric shadow (red region, denoted with $S$) that
cannot be reached by any trajectory, except for normal incidence ($\theta_i = 0^\circ$).
This region covers an area of length
\begin{equation}
 \ell = \frac{\left( 1 - \cos \theta_i \right)}{\cos \theta_i}\
  \Big[ \, a - z_0 \tan \theta_i/2 \ \Big] ,
 \label{eq:2.2}
\end{equation}
which depends on $\theta_i$, vanishing ($\ell = 0$) for $\theta_i = 0^\circ$ and reaching
its maximum extension ($\ell = \infty$) for parallel incidence ($\theta_i = \pi/2$).

\vspace{-3pt}

\begin{figure}[H]
 \centering
 \includegraphics[width=8cm]{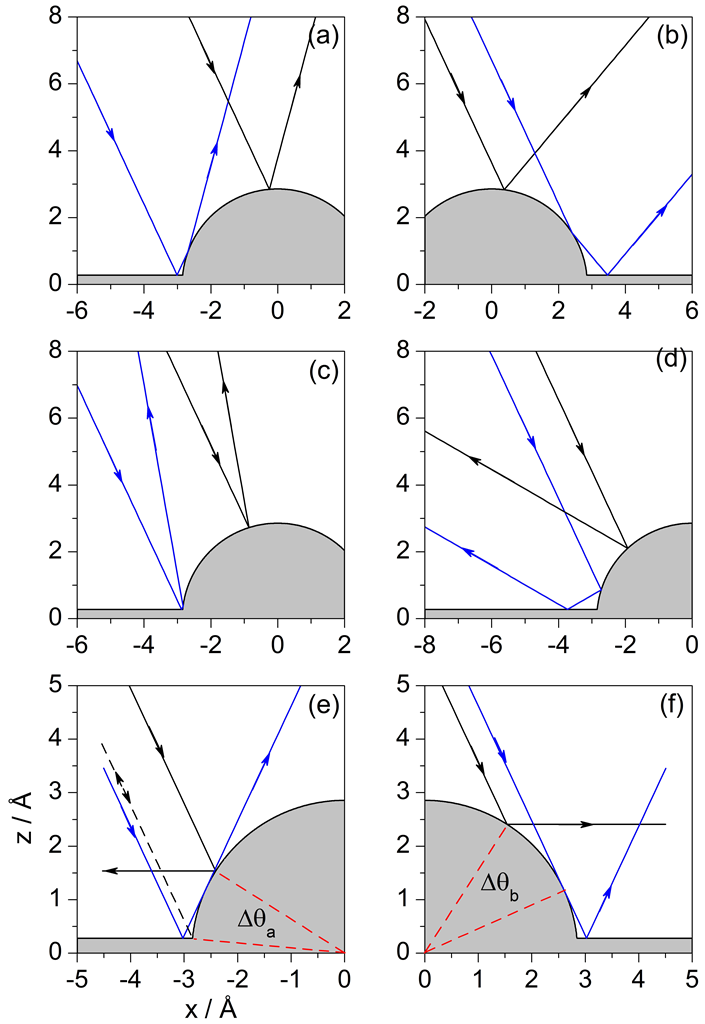}
 \vspace{-6pt}
 \caption{In panels (\textbf{a}--\textbf{d}), pairs of homologous Fermatian trajectories with different deflection angle $\theta_d$:
  (\textbf{a}) forward deflection, with $\theta_i \ge \theta_d \ge 0$; (\textbf{b}) forward deflection, with $\pi/2 \ge \theta_d \ge \theta_i$;
  (\textbf{c})~backward deflection, with $- \theta_i + \delta \le \theta_d \le 0$, where $\delta = (\sin)^{-1} (z_r/a)$ and
  (\textbf{d}) backward deflection, \mbox{$- \pi/2 \le \theta_d \le - \theta_i$.}
  In panels (\textbf{e},\textbf{f}), boundaries of the regions where any trajectory impinging on them will display double collisions (see text for
  further details).}
 \label{Fig6}
\end{figure}

Each trajectory in Figure~\ref{Fig5} carries a label, which helps to identify behavioral
domains.
We have that any trajectory impinging on the substrate either to the left of $\mathcal{F}_\alpha$ or
to the right of $\mathcal{F}_\beta$ will undergo forward deflection; any other trajectory confined in between
will be deflected backwards.
Notice that $\mathcal{F}_\alpha$ and $\mathcal{F}_\beta$ are the only two trajectories that are deflected along
the normal, constituting themselves a pair of homologous trajectories, with $\mathcal{F}_\alpha$ displaying
double collision (first with the flat surface and then with the adsorbate) and $\mathcal{F}_\beta$ displaying
a single collision (with the adsorbate).
As for the other~separatrices:
\begin{itemize}[leftmargin=*,labelsep=5.5mm]
 \item Trajectories to the left of $\mathcal{F}_1$ or to the right of $\mathcal{F}_7$ only interact
  with the clean Pt surface and hence their deflection and incidence angles are equal.
  These trajectories, plus $\mathcal{F}_5$ only contribute to mirror reflection from the flat surface, only
  contributing the intensity for $\Delta K = 0$, since $\theta_d = \theta_i$---hence, this contribution
  will be more prominent as the range of impact parameters~increases.

 \item Any trajectory between $\mathcal{F}_1$ and $\mathcal{F}_\alpha$ is deflected in an angle that
  goes from $\theta_i$ to $0^\circ$ as the impact parameter increases.
  The same deflection angles are found for trajectories between $\mathcal{F}_\beta$ and $\mathcal{F}_5$, although
  here the trend is that the angle increases from $0^\circ$ to $\theta_i$ as $b$ increases.
  Here, we have two sets of pairs of homologous trajectories: trajectories from the first set undergo double collisions
  (first with the flat surface and then with the adsorbate) and trajectories from the latter only have a single
  collision (with the flat surface).
  For any of these pairs, the angular distance between their impact points on the adsorbate
  surface is $\pi/2 - \theta_i$, as can be seen in Figure~\ref{Fig6}a.

 \item There are also pairs of homologous trajectories with deflection angles between $\theta_i$ and $\pi/2$.
  These are the trajectories confined between $\mathcal{F}_5$ and $\mathcal{F}_6$, with single collisions (with
  the adsorbate), and~between $\mathcal{F}_6$ and $\mathcal{F}_7$, with double collisions (first with the adsorbate
  and then with the flat surface).
  This second set corresponds to trajectories impinging on the adsorbate within the sector $B$.
  In this case, the angular distance between impact points is not a constant, but depends on the deflection angle
  as $\pi/2 - \theta_d$.
  This distance gradually vanishes as both trajectories approach $\mathcal{R}_6$ and is maximum when the trajectories
  coincide with the separatrices $\mathcal{R}_5$ and $\mathcal{R}_7$.
  A representative set is depicted in Figure~\ref{Fig6}b.

 \item Trajectories $\mathcal{F}_2$ and $\mathcal{F}_4$ are both deflected backwards along the incidence
  direction, i.e., $\theta_d = - \theta_i$.
  Accordingly, trajectories between $\mathcal{F}_\alpha$ and $\mathcal{F}_2$ are deflected between $0^\circ$ and
  $-\theta_i$ after undergoing double collisions (first with the flat surface and then with the adsorbate), while trajectories
  between $\mathcal{F}_4$ and $\mathcal{F}_\beta$ (to the right of $\mathcal{F}_4$) undergo single collisions.
  The angular distance between impact points of homologous pairs of trajectories is now $\pi/2 - \theta_d$, although not all
  trajectories between $\mathcal{F}_4$ and $\mathcal{F}_\beta$ have a correspondent between $\mathcal{F}_\alpha$
  and $\mathcal{F}_2$.
  This is because the flat surface intersects the adsorbate surface at a distance $z=z_r$ above its center of mass instead
  of at $z=0$.
  Thus, instead of reaching a maximum deflection of $-\theta_i$, we have $\theta_d^{\rm max} = - \theta_i + (\sin)^{-1} (z_r/a)$,
   which is the deflection for the trajectory $\mathcal{F}'_2$.
  An illustrative pair of homologous trajectories of this kind is displayed in Figure~\ref{Fig6}c.

 \item The trajectory $\mathcal{F}_3$ separates the sets of homologous trajectories that are backward deflected, with
  the second collision taking place from the flat surface.
  One set is confined within trajectories $\mathcal{F}_2$ and $\mathcal{F}_3$, with double collisions (first with the adsorbate
  and then with the flat surface), and the other set, with single collisions, is delimited by $\mathcal{F}_3$ and $\mathcal{F}_4$
  (trajectories to the left of $\mathcal{F}_4$).
  Unlike the previous set of backward-scattered homologous pairs, here all trajectories are paired, with the angular distance
  between their impact points being $\pi/2 - \theta_d$, as before.
  A representative pair is displayed in Figure~\ref{Fig6}d.
\end{itemize}
\vspace{-6pt}
\vspace{-6pt}

In general, independently of whether the above pairs of homologous Fermatian trajectories
describe situations of forward or backward scattering, and also regardless of the incidence
angle, double collisions arise whenever the impact on the adsorbate surface takes place
between the point where this surface intersects the flat Pt surface and the point at which
any impinging trajectory is deflected parallel to such a flat surface (at a given incidence).
These conditions determine the regions labeled with $A$, $B$ and $C$ in Figure~\ref{Fig5}.
This is seen with more detail in Figure~\ref{Fig6}e, where the neighboring regions $A$
and $C$ are determined by the separatrices $\mathcal{F}_1$ and $\mathcal{F}_3$,
and in Figure~\ref{Fig6}f for region $B$, delimited by the separatrices $\mathcal{F}_6$
and $\mathcal{F}_7$.
Although regions $A$ and $C$ are close to each other, the double collision process is
reversed when the impact parameter passes from the domain of one of them to the other.
Thus, we find that while in regions $A$ and $B$ the atom first collides with the adsorbate
and then with the flat surface, in the case of region $C$ it is the opposite.
The length of this latter region is nearly the same as the one for the shadow
region $S$, given by Equation~(\ref{eq:2.2}), because of the symmetry between the trajectories
$\mathcal{F}_1$ and $\mathcal{F}_7$
On the other hand, regions $A$ and $B$ are defined, respectively, by the
angular sectors $\Delta \theta_A = \pi/4 - \theta_i/2 - (\sin)^{-1} (z_0/a)$ and
$\Delta \theta_B = \pi/4 - \theta_i/2$, which are both dependent on the incidence angle.

Regarding trapping, although this simple potential function cannot lead to this phenomenon,
it is interesting to note that, to some extent, $\mathcal{F}_3$ and $\mathcal{F}_7$ can be
considered as permanently trapped trajectories, since they will keep evolving parallel to the
surface ($\theta_d = \pm \pi/2$).
This is, of course, a~rather weak case associated with a purely repulsive model, but still it
is useful to understand in simple terms the types of dynamics that can be expected from a
more refined classical model, such as the one specified by (\ref{eq:1.1}), concerning the
presence of homologous pairs of trajectories (associated with single and double collisions)
as well as the appearance of trapping.


\subsection{Newtonian Level}
\label{sec42}

The hard-wall model constitutes a sort of crude approach to the system studied,
appropriate to explain some general features.
However, a more realistic or refined model (closer to the experimental system) is
going to include additional features, such as the rainbow phenomenon, which have also
received much attention both experimentally \cite{lahee:JCP:1987,graham:JCP:1996} and theoretically
\cite{yinnon:JCP:1988,lemoine:JCP:1994-1,lemoine:PRL:1998}, or the defect-mediated diffraction resonance
\cite{yinnon:JCP:1988,glebov:PRL:1997}, a kind of trapping induced by the presence of adsorbates on
surfaces.
These phenomena are associated to the particularities displayed by the potential
model, such as the depth of potential wells, the stiffness of the repulsive region
or the range of its attractive region.
All these features are controlled by means of parameters that are found from best
fit to the cross sections (intensities) experimentally measured; the optimal values
eventually provide us with valuable information on the system analyze.
In the particular case of the system discussed in this work, such information has to
do with the way how the CO is attached to Pt(111) surface, which can be later used
to better understand reactivity and diffusion properties.

\begin{figure}[H]
 \centering
 \includegraphics[height=9cm]{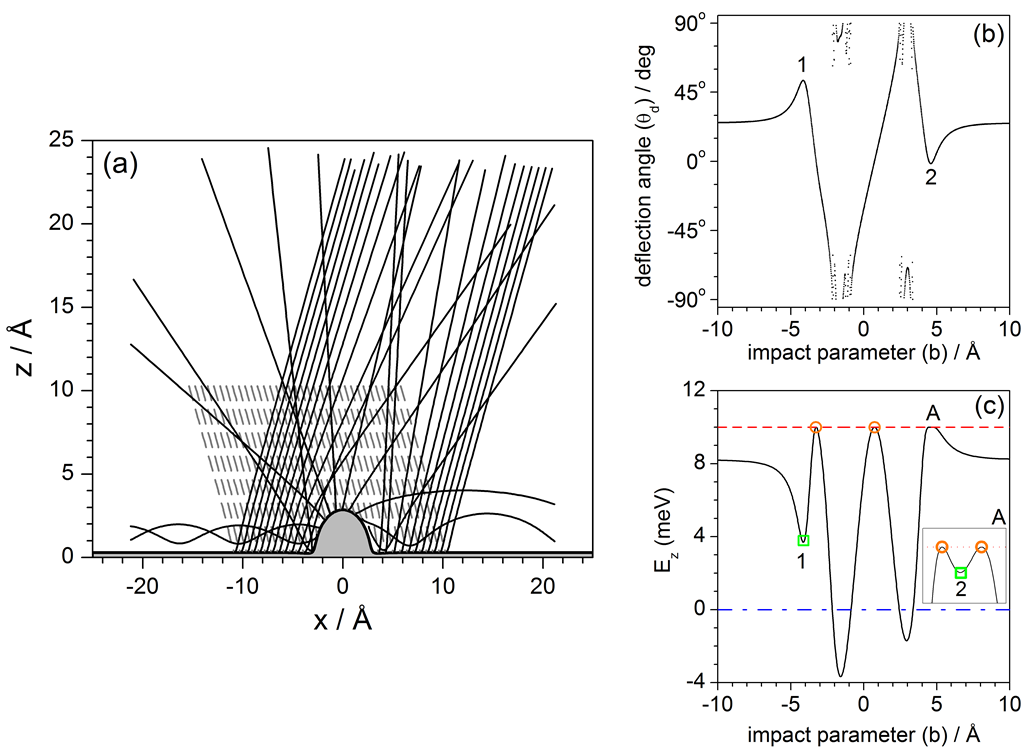}
 \vspace{-6pt}
 \caption{(\textbf{a}) Set of classical Newtonian trajectories for the interaction potential model (\ref{eq:1.1})
  for an incidence energy $E_i = 10$~meV.
  As in Figure~\ref{Fig5}, for a better illustration of all possible cases, also an incidence angle $\theta_i = 20^\circ$
  has been chosen.
  Moreover, the incident part of the trajectories has been represented with dashed line;
  (\textbf{b}) classical deflection function.
  While surface trapping gives rise to two kind of discontinuous regions, rainbow features manifest as local
  maxima (1) and minima (2);
  (\textbf{c})~Asymptotic-energy diagram.
  Here, trapping is detected through the two regions of the curve below the threshold $E_z = 0$~meV, while
  rainbows manifest with two local minima (green squares; for rainbow~2, see enlargement of region A in the inset).
  Orange circles denote conditions leading to perpendicular deflection with respect to the flat surface.}
 \label{Fig7}
\end{figure}

The dynamics under the influence of the potential model (\ref{eq:1.1}) are illustrated
in Figure~\ref{Fig7}a by means of a set of trajectories with initial conditions uniformly
covering a wide range of impact parameters.
Specifically, in this simulations $b \in [-10.6,10.6]$~\AA, for $E_i = 10$~meV and $\theta_i = 20^\circ$
(again, as in the previous section, incidence out of the diagonal has also been chosen for these trajectories
to stress some particular aspects).
The attractive long-range term from Lennard--Jones contribution plays an important role
here, since it is responsible for the permanent trapping (in this model) of He atoms
along the Pt surface.
This term accounts for the van der Waals interaction mediating between the neutral He
atoms and the CO adsorbate, producing an effective transfer of energy from the
perpendicular direction to the parallel one, such that the energy along this latter
direction becomes larger than the incident energy to the expense of making negative
the energy along the $z$ direction.
Of course, surface trapping is not totally permanent, since the presence of other
adsorbates (not considered here, where we are working under single-adsorbate
conditions) produces the opposite effect, that is, a trapped atom, after colliding
with another neighboring adsorbates, may acquire enough energy along the $z$-direction
(loosing it along the $x$ direction) to escape from the surface.

\vspace{-6pt}

A useful tool to systematize and analyze the different dynamical behaviors exhibited
by the trajectories of Figure~\ref{Fig7}a is the deflection function, i.e., the
representation of the deflection angle with which the He atoms asymptotically leave
the substrate as a function of their impact parameter.
This~function is represented in Figure~\ref{Fig7}b and clearly shows that it is
characterized by two types of regions.
One of them is smooth and continuous, which means that the deflection angle increases
or decreases gradually.
The local maxima and minima within this type of region denote the presence of rainbows,
i.e., deflection directions characterized by an extremely high intensity (leaving aside
mirror reflection from the flat surface).
In the figure, we observe the presence of two of these rainbows (denoted with the numbers
1 and 2), one of them nearly along the normal to the flat Pt surface.
The other type of region is seemingly random, which is a signature of trapping---this makes
the function in these regions to be time-dependent, since deflection will depend on the time
at which it is computed.
Nonetheless, this behavior is not to be misinterpreted with presence of chaos that we observe
in analogous representations for He diffraction from corrugated surfaces \cite{sanz:SSR:2004}.
The difference is that, while, in those cases, this random-like region has a fractal structure
\cite{sanz:SSR:2004}; here, it is very regular.
This can easily be seen by plotting more values of the impact parameter within this region.
Then, as this region of the deflection function becomes more dense, we will be able to
better appreciate its regularity and, therefore, the lack of an underlying chaotic dynamics.

To complement as well as to disambiguate the information provided the by
deflection function, particularly in the trapping regions, an alternative
representation can be obtained if we focus far from the adsorbate, where
the system energy is separable, as mentioned above.
Accordingly, unless there is an extra energy exchange because of the presence
of other defects, the He atom will keep constant its energies along the $x$
and the $z$ directions (there is no coupling term between both degrees of freedom
in the potential describing the flat Pt surface).
This thus allows to consider energy diagrams, where the energy left along the $z$
degree of freedom is compared to the total available energy, which coincides with
the incident energy ($E_i$), and to the dissociation threshold (0~meV), which
determines the minimum energy for unbound motion.
This plot is shown in Figure~\ref{Fig7}c for $E_i=10$~meV
In this representation, we first note that a large portion of impact parameters gives
rise to unbound motion, i.e., all the trajectories leave the Pt surface.
Among these trajectories, those leaving the surface at an angle equal to a rainbow
angle produce a local minimum in the energy diagram (green squares)---the absolute value
of the rainbow angle can then be readily determined by means of the simple relation
$\cos \theta_R = \sqrt{E_z^R/E_i}$---, while maxima
(orange circles) correspond to trajectories that leave the surface perpendicularly (no
energy along the $x$ degree of freedom, although initially they all started with some
energy in it).
On the other hand, we also find some impact parameter regions for which the energy
is negative.
These regions correspond to conditions leading to trapping (oscillatory bound motion
parallel to the flat surface), which are in correspondence with the discontinuous
regions observed in the deflection function.
Actually, from the energy diagram, we can determine with high accuracy the precise
impact parameters that, at a certain incidence, will give determine the ranges of
trapping.
From~the figure, we see that these limits are given by the intersection points of
the energy curve with the zero-energy condition.

\begin{figure}[H]
 \centering
 \includegraphics[height=8cm]{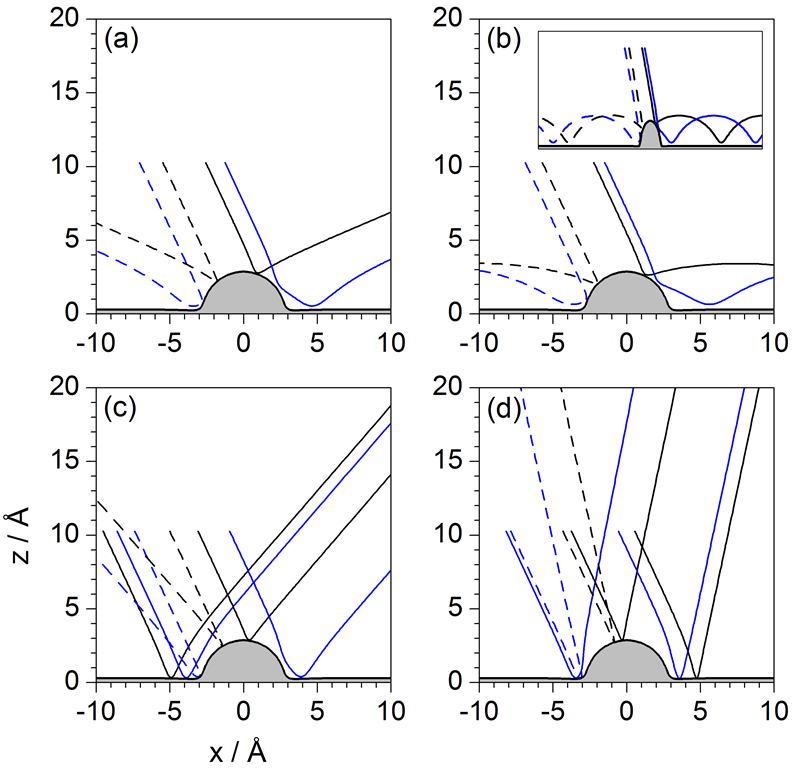}
 \caption{Pairs of homologous Newtonian trajectories with the same asymptotic value for their energy along the
  $z$, obtained from the energy diagram of Figure~\ref{Fig7}c: (\textbf{a}) $E_1 > E_z > 0$; (\textbf{b}) $0 > E_z$;
  (\textbf{c}) $E_i \cos^2 \theta_i > E_z > E_1$ and (\textbf{d}) $E_2 > E_z > E_i \cos^2 \theta_i$,
  where $E_1$ and $E_2$ are the energies (along the $z$ direction) corresponding to rainbows 1 and 2
  (see Figure~\ref{Fig7}b,c),
  respectively, and $E_i \cos^2 \theta_i$ is the incidence energy (also along the $z$ direction), with $E_i = 10$~meV.
  Forward/backward deflected trajectories are denoted with solid/dashed line.
  Trajectories undergoing single/double collision/s are denoted with black/blue colors.
  All trajectories are started from a distance $z_i = 10.27$~\AA\ above the flat Pt surface
  (beyond $z \approx 6.35$~\AA, the interaction potential model (\ref{eq:1.1} is negligible;
  see Figure~\ref{Fig1}a).}
 \label{Fig8}
\end{figure}

As we have seen, the energy diagram and the deflection function, in general terms, provide
the same kind of information.
Now, the energy diagram can be smartly used to determine pairs of homologous Newtonian
trajectories as follows.
Consider a given value for $E_z$.
All the impact parameters that are obtained from the intersection of a horizontal line at
the selected value of $E_z$ with the energy diagram will provide sets of pairs of homologous
trajectories.
Some illustrative pairs are shown in Figure~\ref{Fig8}.
In panel (a), for a selected energy such that $E_1 > E_z > 0$, where $E_1$ is the energy (along
the $z$ direction) for rainbow 1, we find two pairs of trajectories, one back-scattered and another
forward-scattered.
Although distinguishing between single and double collisions is not as simple as with the hard-wall
model, we still can perceive that within these two pairs of homologous trajectories one of them
undergoes a single collision (denoted with black color), while the other shows something that could
be related with a double collision (with blue color).
If $E_z$ is below zero, we can still have two pairs of homologous trajectories, as seen in panel (b),
although these trajectories exhibit permanent trapping.
For energies above $E_1$ we can observe up to three or (for $E_z > E_2$) four pairs of homologous
trajectories, with analogous behaviors.
Actually, unlike the hard-wall model, here we notice that there can be several pairs contributing
to reflection symmetry interference, as happens in panels (c) and (d).

\begin{figure}[H]
 \centering
 \includegraphics[height=6cm]{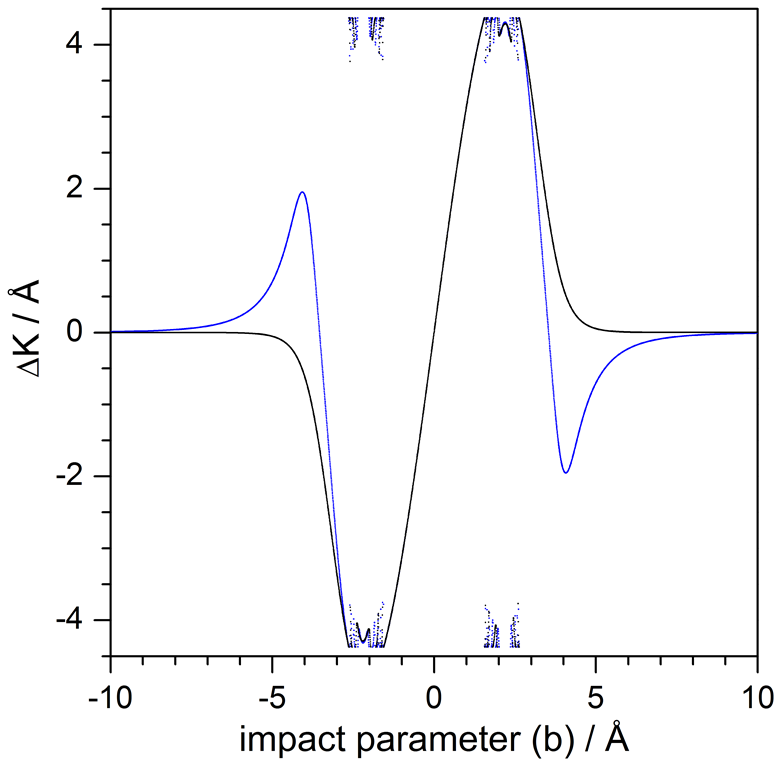}
 \caption{Deflection function for a purely repulsive model of the CO adsorbate (black solid line), obtained after
  removal of the attractive part of the Lennard--Jones function (\ref{eq:1.3}).
  Here, for simplicity, the incidence conditions are $\theta_i = 0^\circ$ and $E_i = 10$~meV.
  To compare with, the deflection function corresponding to the full interaction potential model (\ref{eq:1.1}) is
  also represented (blue solid line), obtained for the same incidence conditions.}
 \label{Fig9}
\end{figure}

Finally, Figure~\ref{Fig9} offers a comparison between the behavior displayed by a
purely repulsive adsorbate (closer to the hard-wall model) and the attractive
adsorbate in terms of the corresponding deflection functions for $E_i = 10$~meV and
normal incidence.
The main difference between both models relies on a removal of the attractive term
of the Lennard--Jones model.
Notice that the repulsive model lacks the two rainbow features for $\Delta K_R = \pm 1.95$~\AA$^{-1}$
($\theta_R = \pm 26.50^\circ$), although both models keep nearly the same trapping rates, which has to
do with the effective energy transfer from the perpendicular to the parallel directions due to the
adsorbate curvature, and not that much with the presence of attractive basins around the adsorbate
itself.
Although not shown here, the same holds if the energy is increased.
For instance, for $E_i = 40$~meV and normal incidence, rainbows are observed for $\Delta K_R = \pm 1.21$~\AA$^{-1}$
($\theta_R = \pm 7.96^\circ$) with the full potential model (\ref{eq:1.1}), but there is a complete absence of them
when the repulsive model is used instead.

\vspace{-12pt}

\subsection{Bohmian Level}
\label{sec43}

In the literature there has always been a controversy concerning how the
different diffraction features observed in intensity patterns should be
assigned in the case of the atom diffraction by impurities on surfaces.
In 1988, for instance, Yinnon et al.~\cite{yinnon:JCP:1988} suggested the use of
the quantum flux, ${\bf J}$, as an interpretational tool to understand
trapping processes in this type of systems.
A vector representation of this field provides a reliable representation of
how the Fraunhofer, rainbow and trapping features arise, although is not
unambiguous at all.
This procedure was employed earlier on in reactive scattering by Wyatt
\cite{mccullough:JCP:1969,mccullough:JCP:1971-1,mccullough:JCP:1971-2}.
Although some dynamical information can be extracted about the distribution
and change of the flux, it is difficult to understand how the system
eventually evolves.
The last step in our journey is precisely the use of Bohmian trajectories
to get an idea of what is going on this type of systems from at a fully
quantum level and, beyond quantum flux based analyses, to~understand the
dynamics by causally connecting a point on the initial state with another
point of the final state, following a well-defined trajectory in real
time.
This trajectory is obtained by integration of the equation of motion
\cite{sanz-bk-1,sanz:EJP-arxiv:2017}
\be
 \dot{\bf r} = \frac{{\bf J}}{\rho}
  = \frac{\hbar}{2mi} \left( \frac{\Psi^* \nabla \Psi - \Psi \nabla \Psi^*}{\Psi \Psi^*} \right)
  = \frac{\nabla S}{m} ,
 \label{eqbohm}
\ee
often regarded as guidance equation.
In this equation, formerly introduced by Bohm as a postulate~\cite{bohm:PR:1952-1,holland-bk}, $\rho$ and
${\bf J}$ are respectively the probability density and the quantum flux \cite{schiff-bk}; the velocity
field ${\bf v} = \dot{\bf r}$ is just the way how the probability density spreads through the configuration
space in the form of a flux or current density.
Regarding $S$, it is the phase field, which describes the local variations of the quantum phase and is
typically obtained from the polar transformation of the wave function,
\be
 \Psi ({\bf r},t) = \sqrt{\rho ({\bf r},t)} e^{i S({\bf r},t)/\hbar} .
\ee
The Bohmian trajectories shown below are obtained taking into account the value of the wave function at
a given time (obtained by means of the propagator mentioned in Section~\ref{sec2}), according to the second
expression for the velocity field in Equation~(\ref{eqbohm}).

\begin{figure}[H]
 \centering
 \includegraphics[width=11cm]{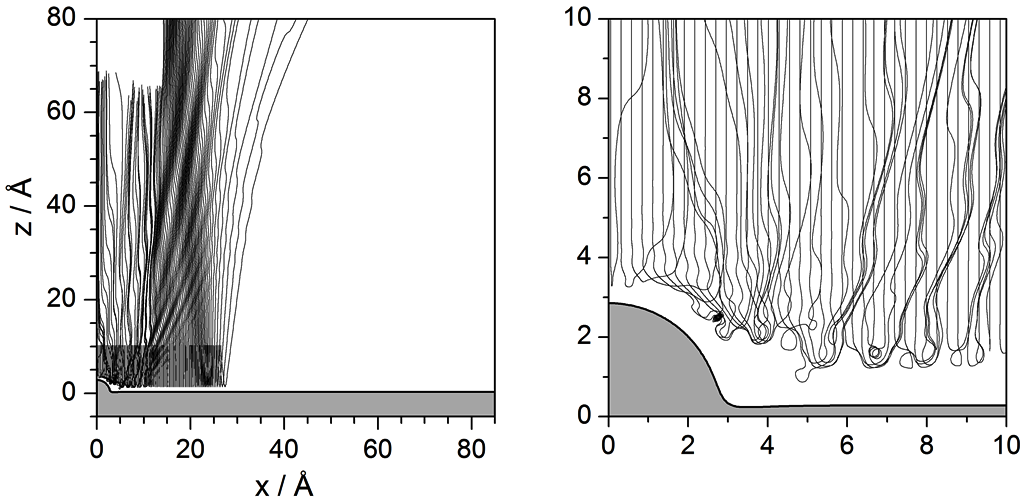}
 \caption{Set of Bohmian trajectories with initial positions uniformly distributed along
  the $x$ direction and fixed value along the $z$ direction: $z_i = \langle z \rangle_{0} = 10.27$~\AA,
  which corresponds to the center of the incident wave packet ($t=0$) above the clean Pt surface
  [beyond $z \approx 6.35$~\AA, the interaction potential model (\ref{eq:1.1}) is negligible;
  see Figure~\ref{Fig1}a].
  On the right-hand side, enlargement of the left plot near the adsorbate to illustrate the dynamical
  behavior displayed by the trajectories close to the substrate~surface.}
 \label{Fig10}
\end{figure}

\begin{figure}[H]
 \centering
 \includegraphics[width=12cm]{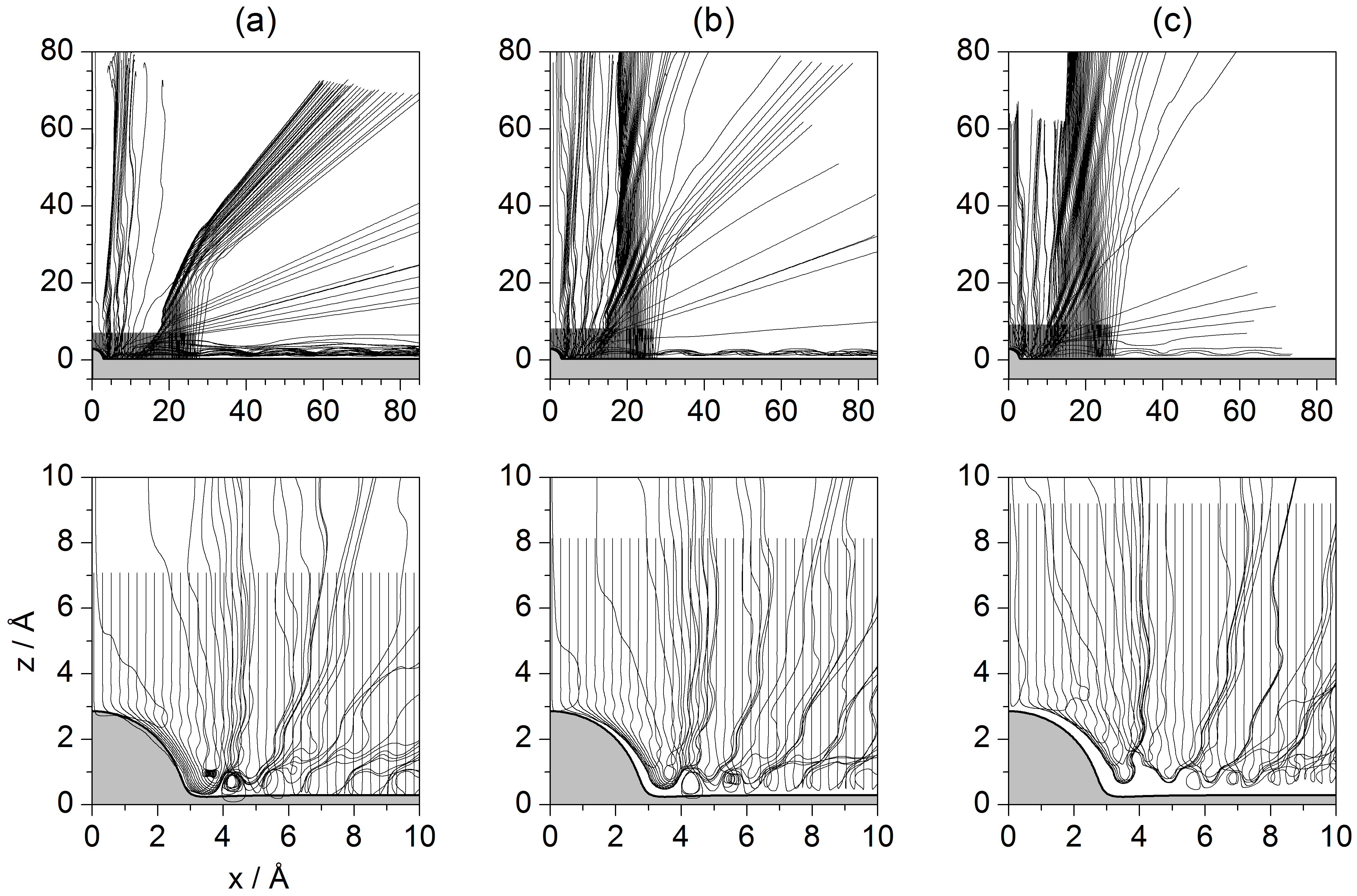}
 \vspace{-6pt}
 \caption{Set of Bohmian trajectories with initial positions uniformly distributed along
  the $x$ direction and fixed value along the $z$ direction:
  (\textbf{a}) $z_0 = \langle z \rangle_{0} - 3.18$~\AA; (\textbf{b}) $z_0 = \langle z \rangle_{0} - 2.12$~\AA~and
 \mbox{ (\textbf{c}) $z_0 = \langle z \rangle_{0} - 1.06$~\AA,} where $\langle z \rangle_{0} = 10.27$~\AA, which
  corresponds  to the center of the incident wave packet ($t=0$) above the clean Pt surface
  (beyond $z \approx 6.35$~\AA, the interaction potential model (\ref{eq:1.1}) is negligible;
  see Figure~\ref{Fig1}a).
  In the corresponding lower panels, enlargement of the upper plots near the adsorbate to illustrate the
  dynamical behavior displayed by the trajectories close to the substrate~surface.}
 \label{Fig11}
\end{figure}

\begin{figure}[H]
 \centering
 \includegraphics[width=12cm]{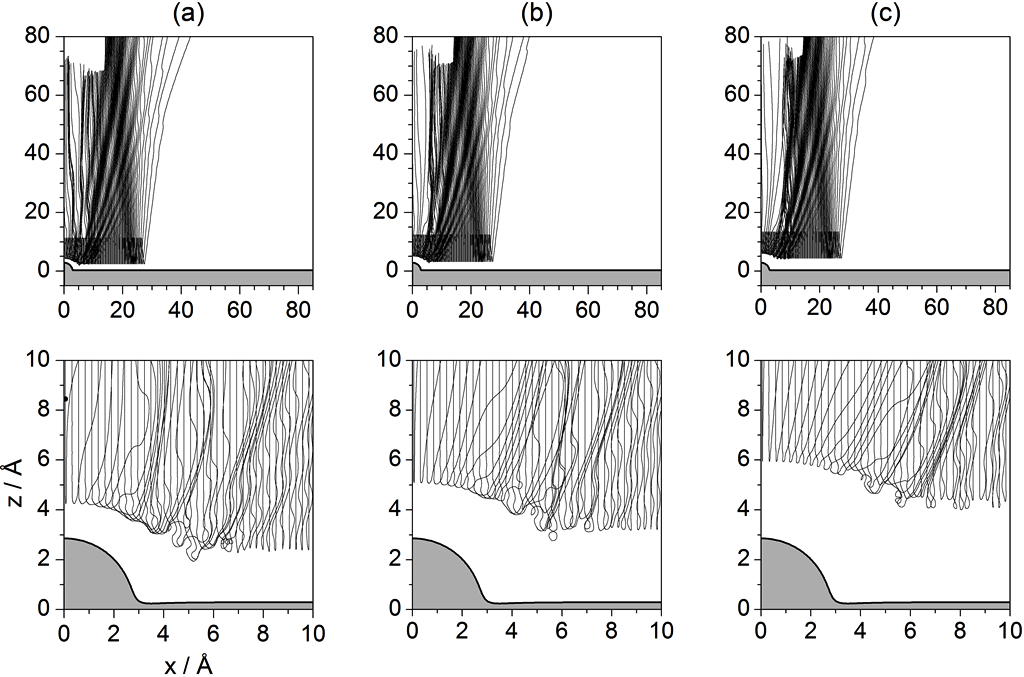}
 \vspace{-12pt}
 \caption{As in Figure~\ref{Fig11}, but considering:
  (\textbf{a}) $z_0 = \langle z \rangle_{0} + 1.06$~\AA; (\textbf{b}) $z_0 = \langle z \rangle_{0} + 2.12$~\AA~and
  (\textbf{c})~$z_0 = \langle z \rangle_{0} + 3.18$~\AA.}
 \label{Fig12}
\end{figure}

The analysis in this section is performed by studying the behavior displayed
by sets of Bohmian trajectories with initial positions taken at a series of
distances from the surface and uniformly distributed along the parallel
direction for each one of those distances.
Unlike the procedure followed in preceding sections, now the loss of translational
symmetry caused by the presence of the isolated adsorbate on the Pt surface does
not allow for studying the dynamics considering impact parameters only along the
parallel direction for a given $z$ value.
This is the reason why different values of $z$ are considered.
This enables a better way to assign the different parts of the incident wave
with final outgoing intensity peaks without ambiguity.
These sets are displayed in Figures~\ref{Fig10}--\ref{Fig12}, with each
set being labeled with the corresponding initial condition along the $z$ direction
referred to the center of the wave packet $\langle z \rangle_0 = 10.27$~\AA.
According to these figures, quantum-mechanically the dynamics cannot be understood
in the same local terms as in classical mechanics, where trajectories did not
display a different behavior depending on their starting distance along the vertical
direction with respect to the adsorbate.
In order to obtain a complete knowledge of the diffraction process, trajectories
have to be chosen from across the whole region covered by the initial wave function.
Different regions will give rise to different diffraction features, but also different
sets of trajectories will be able to probe the surface at a different level, being
able to approach it very closely or, on the contrary, will bounce backwards far
away, without even having touched it physically.
In this regard, one wonders whether, contrary to what is commonly stated within the ``Bohmian
community'', the (Bohmian) trajectories can be considered to reveal the ``true'' motion followed
by the particles they are associated with---in the present case, the motion of individual
He atoms.
This is a rather challenging question (as well as metaphysical), which, to the author's best
knowledge, has not still been unambiguously answered, that is, with a solid, irrefutable
experimental proof.
Therefore, taking on a pragmatic view, here Bohmian trajectories are considered as hydrodynamic
streamlines that allow us to investigate the flow dynamics of the probability flux in configuration
space and, therefore, to understand towards which directions atoms are more likely redirected
(deflected) after being scattered from the substrate (without providing any particular information
on how each individual atom really moves).
Actually, to some extent, this random view of the atomic motion is the idea behind the work that
Bohm developed in 1954 in collaboration with Vigier \cite{bohm:pr:1954} and, later on, in 1989 with
Hiley \cite{bohm:PhysRep:1989}, or the approach developed in 1966 by Nelson \cite{nelson:pr:1966}
(although this latter approach has nothing to do with Bohmian mechanics, it introduces analogous
stochastic concepts).
In these examples, the quantum particle follows a random-like path as a consequence of the action
of a sub-quantum random medium; when motions are averaged, particle statistics reproduce the
results described by Schr\"odinger's equation and Bohmian trajectories arise as the averaged
flow-lines associated with the solutions to this equation.

\vspace{-12pt}

In Figures~\ref{Fig11} and \ref{Fig12}, sets of trajectories taken from below and from
above the $z$-value selected in Figure~\ref{Fig10} are shown.
In the three cases displayed in Figure~\ref{Fig11}, we notice that those trajectories
that start closer to the surface are going to contribute more importantly to
the marginal portions of the outgoing wave ($B_i$ and $C$, as defined in Figure~\ref{Fig3}d).
In sharp contrast, trajectories started in regions far from the adsorbate (see
Figure~\ref{Fig12}) will contribute to the peaks related to small $\Delta K$ ($A_i$).
This can be understood establishing a nice analogy with classical fluid dynamics and
then introducing the notion of quantum pressure introduced by Takabayashi \cite{takabayasi:ProgTheorPhys:1952}.
Accordingly, when the wave function is on the surface, we find that, while its lowest
part is already bouncing backwards, the upper one is still moving downwards (towards
the surface).
In this situation, in the analogy of the wave function being associated with an ideal
non-viscid and incompressible fluid, its upper part would be pushing the lowest one
and then generating a remarkable pressure on it (this effect would increase with the
incidence energy, although its duration would be shorter).
This is something that cannot be seen directly on a wave-function representation, but
that has an interesting counterpart in the case of the trajectories because we can see
that those associated with the lowest part of the incident wave function are then pushed
(or make evident the push) against the surface, remaining there for a rather long time.
On the contrary, as the initial conditions are taken from upper regions of the initial
wave function, the pressure on them will be lower and hence will bounce backwards
further away from the physical surface.
Thus, the evolution of these trajectories resembles in a closer way the behavior of classical
trajectories, while the trajectories below are somehow forced to propagate parallel
to the surface and escape by the borders of the wave.

The turbulent dynamics manifested by the trajectories close to the surface has
to do with the web of nodal lines characterizing the wave function when it is on the
surface.
In particular, it is interesting that how sometimes the trajectories get trapped
and whirl around some of these nodes, undergoing a sort of transient vortical
trapping \cite{sanz:prb:2004,sanz:jcp:2004}, different from the temporary trapping of the trajectories
that will remain confined within the surface attractive well far from the adsorbate.
The whirlpool motion displayed by such trajectories has actually an interesting
property, namely the associated action is quantized, being an integer multiple
(equal to the number of full loops) of a certain value.
This~type of motion is a confirmation of the former results found by
Yinnon et al.~\cite{yinnon:JCP:1988}.
Actually, recently, Efthymiopoulos {et al.}~\cite{efthymio:AnnPhys:2012} formally determined
the conditions under which such temporary nodal regions in scattering problems will appear, which
happens to be in the regions where the amplitudes (modulus) of the incoming and outgoing waves
become equal.
This boundary, where both values are the same and that have important dynamical consequences,
is what they called ``separator''.

Regarding the reflection symmetry interference phenomenon, this time it is not as simple
as in the two previous cases because the trajectories obey the guiding rules imposed by
the wave function and not the direct action of the potential.
In the case of the rainbow, however, there is a more evident manifestation.
Although Bohmian trajectories cannot cross one another at the same time, collectively they
show a sort of rainbow-like precession, which is more prominent in the case of the trajectories
shown in Figure~\ref{Fig11}:
after the collision with the adsorbate, they start showing larger and larger deflections until
getting trapped, then the start precessing backwards until reaching a maximum angle, and finally
precess backwards again to smaller angles.
In the case of the trajectories displayed in Figure~\ref{Fig12}, the behavior is analogous,
even if in this case there is no trapping and the effect is more subtle.
Thus, all the ensembles show a similar behavior, supporting the idea of the rainbow as
being a sort of ``global'' effect, which translates into the ``wings'' observed in the
intensity pattern shown in~Figure~\ref{Fig4}.


\section{Conclusions}
\label{sec5}

In the last several years, Bohmian mechanics has been gaining ground as an appealing tool to deal with
very different problems out of the area of the quantum foundations, its traditional environment.
In such cases, the interest and relevance of Bohmian mechanics is emphasized by directly
tackling a given problem with it.
In the current work, however, the motivation has been a bit different.
There are different trajectory-based approaches that have been or can be used to describe,
analyze, understand and explain quantum systems and phenomena, even if they have
different degrees of accuracy. The~purpose here has been to establish an appropriate context to better understand the role
of Bohmian trajectories within all those formulations as well as the kind of information conveyed
by each one at its level of accuracy.
To this end, we have considered a problem of interest out of the field of the quantum
foundations, specifically the diffraction of helium atoms by a nearly flat platinum surface
on top of which there is a carbon monoxide adsorbate.
This is a problem that has been considered in the literature due to its intrinsic practical
applications, although here it has been chosen due to its suitability to the purpose, since
it is sufficiently simple to allow us its treatment at different levels.

Accordingly, the system has first been analyzed with a usual wave-function-based
framework, investigating the effects associated with two different He-CO/Pt(111) interaction
potential models:
\begin{itemize}[leftmargin=*,labelsep=5.5mm]
 \item {\bf Hard-wall model.} This model is in the form of an impenetrable (fully repulsive) wall,
  where the interaction is reduced to a sudden impact on the He atoms on the such a wall.
  The first model allows an exact asymptotic analytical treatment, convenient to elucidate the
  main mechanism observed in the diffraction pattern produced by single adsorbed particles on
  nearly flat surfaces, namely reflection symmetry interference.

 \item {\bf Potential energy function.} This interaction model is determined from fitting to the
  experimental data and constitutes a refinement of the previous one in the sense that there is
  detailed information on the intensity of the interaction between the incoming atom and the
  substrate at each point (in this regard, the hard-wall model is just a crude approximation).
  Thus, in spite of its lack of analyticity, unlike the hard-wall model, it provides us with a more realistic
  description of the diffraction process in real time, rendering information on additional physics,
  such as rainbow features or surface trapping.
\end{itemize}
Although at a different degree of accuracy, these two models provide us with explanation of the
features observed in the experimental diffraction patterns.
The question is how to interpret or understand these diffraction features or, in other words, to elucidate
the physical mechanism responsible for each of such features (reflection symmetry interference,
rainbows, or surface trapping).
Typically, this is done by setting protocols based on quantum-classical correspondence, e.g.,
analyzing the system by means of classical trajectories and then comparing the results rendered
by both the classical (trajectory) model and the quantum-mechanical (wave function) one.

Bearing that in mind, as has been seen in the preceding sections, here we have tackled the issue at three
different levels:
\begin{itemize}[leftmargin=*,labelsep=5.5mm]
 \item {\bf Fermatian level.} This first level is the simplest one, based on computing what has been here
  denoted as Fermatian trajectories, which are just the direct analog to optical rays reflected on a hard wall in a medium with constant refractive index.
  According to this trajectory model:
    \begin{itemize}
   \item These trajectories have revealed that there are pairs of homologous trajectories, such that one of the
    peers undergoes single scattering off the interaction potential, while the other undergoes double scattering.
    The fact that a trajectory collides with the CO/Pt system at one point (single collision) or at two different
    points (double collision) is a function of the impact parameter.
    Accordingly, a simple mapping can be establish, which helps to easily localize regions of impact parameters
    that are going to produce homologous pairs of trajectories.
\vspace{6pt}

   \item The mechanism of {reflection symmetry interference} is associated with these paired trajectories,
    which is explained in the same way that we explain interference from two coherent sources: interference
    maxima and minima arise depending on whether the path difference between the two paths (or virtual rays)
    joining each source with a given observation point on a distant screen is equal to an integer number of
    wavelengths or to half an integer, respectively.
    Although these paths are nonphysical (they are just a mathematical construct), they allow us to understand in
    simple terms the appearance of the alternating structure of bright and dark interference fringes.
    In the present case, the path length arises from the extra path length of the trajectory affected by the double
    collision with respect to the homologous pair with single collision.
\vspace{6pt}
   \item In addition, it has also been seen that two specific trajectories are deflected {parallel} to the surface,
    which can be interpreted as a mechanism precursor of the { surface trapping} mechanism that appears
    in more refined models, such as the Newtonian and the Bohmian~ones.
  \end{itemize}

\item {\bf Newtonian level.} On the next level, the Newtonian one, classical trajectories are obtained for the
  realistic potential energy surface describing the interaction between the He atoms and the substrate.
  In this case, it is not so simple to distinguish between single and double collisions, because the deflection of the
  trajectories near the surface, where the interaction between the He atoms and the CO/Pt surface is stronger,
  changes gradually very smoothly.
  However, we have been able to extract a series of interesting conclusions:
  \begin{itemize}
   \item By means of an energy diagram (asymptotic energy along the $z$ direction as a function of the impact
    parameter), we been able to devise a method that allows to determine in a simple fashion pairs of homologous
    (Newtonian) trajectories.
    This diagram is thus a suitable method to determine a behavioral mapping of initial conditions (impact parameters)
    for a given incidence direction (incident energy).
\vspace{6pt}
   \item Accordingly, also at this level, it is possible to find an underlying mechanism responsible for the {reflection
    symmetry interference} found in the corresponding quantum intensity patterns.
    Actually, interference patterns could be reconstructed in the same way as with the Fermatian model, although in
    this case we would be dealing with a space-dependent refractive index (the potential function) and the Newtonian
    trajectories would play the role of Feynman's paths.
    Nonetheless, although such a reconstruction is possible and the techniques are well known, this does not mean that
    trajectories, Fermatian or Newtonian, contain any information on the interference process; in both cases, they are
    only a tool to determine the interference pattern.
\vspace{6pt}
   \item Regarding the {trapping phenomenon}, it has been found to be more prominent, with an important amount
    of trajectories remaining trapped permanently along the surface.
    This is, however, only a {temporary} feature, since it may disappear as son as the trapped atoms find another adsorbate.
    In such a case, the collisions with this adsorbate may provoke an effective transfer of energy from the parallel to the
    normal direction, such that the will be able to eventually leave the surface.
\vspace{6pt}
   \item Finally, due to the attractive well surrounding the adsorbate, we have also observed the appearance of {rainbow
    features}, i.e., high accumulations of trajectories along particular deflection directions.
    However, rather than contributing with a specific, localized feature in the corresponding quantum intensity pattern, rainbows
    seem to manifest affecting them globally, i.e., giving rise to features that appear at different places.
    This has been noticed by computing exactly the same with an alternative repulsive adsorbate model, which lacks the
    surrounding attractive well and therefore does not give rise to the formation of rainbows.
 \end{itemize}

 \item {\bf Bohmian level.} The upper level here considered is the Bohmian one, where things change substantially if we note
  that the transition from the Fermatian level to the Newtonian one can be seen as a refinement associated with having a more
  accurate description of the interaction potential model, changing a hard wall by a ``soft'' wall.
  These are the main findings at this level:
\begin{itemize}
   \item First of all, since Bohmian trajectories are associated with a particular wave function, there is no freedom to
    choose a given set of initial conditions because depending on the positions selected relative to the region covered by
    the initial wave function, the trajectories are going to exhibit a different behavior.
    Thus, we have seen that while some of them are deflected quite far from the physical surface (more intense interaction
    region), other trajectories move just on top of it, displaying signatures of {vorticality}.
\vspace{6pt}
   \item To better understand that point, notice that Fermatian trajectories are only ruled by the law of reflection, while
    Newtonian trajectories are ruled by correlations between the two degrees of freedom, $x$ and $y$, that can be locally
    established within the interaction region (i.e., the region where the interaction potential is stronger, near the substrate).
    In the case of Bohmian trajectories, the dynamics is not directly ruled by the interaction potential, but~by a wave field
    that is able to (non-classically) convey information from everywhere in the configuration space (through its phase).
    This makes a substantial difference between classical (Fermatian or Newtonian) and Bohmian trajectories, which
    may lead us to think that direct comparisons or analogies must be taken with care.
    That is, nothing of what has been seen at the previous levels remains at the upper one, since it is not possible to form
    pairs of homologous trajectories.
\vspace{6pt}
   \item In this case, and contrary to the two previous models, the trajectories contain information about the interference
    process and, therefore, can be used to determine the fringe structure of the pattern by simply making statistics over them.
    If they are properly distributed across the region of the configuration space covered by the initial probability density, they
    will eventually distribute according to the final probability distribution by virtue of the continuity equation that they satisfy.
\vspace{6pt}
   \item Regarding rainbow features, present in the Newtonian model and also, with a weak precursor, in the Fermatian one, the
   only a similar behavior is observed, although it is difficult to establish a unique correspondence with the phenomenon of
   the two previous models.
   In the Bohmian case, taken the trajectories that start with the same value $z_0$, it is seen that their final positions show,
   for some range of $x_0$ values, a certain ``precession'' as $x_0$ increases.
   However, it has not been possible to uniquely identify this phenomenon with the classical rainbow.
  In the case of surface trapping, on the contrary, there same effect has been observed in the three models (again, in the
  Fermatian model it is only a weak precursor).
\vspace{6pt}
   \item Finally, it has also been observed that, depending on how close or far a Bohmian trajectory is started from the
    physical substrate surface, it will be able to reach this surface or just bounce backwards quite far from it (from
    what we could call an effective nonphysical surface).
    Actually, if the trajectories start close to the surface, they are influenced by the web of maxima developed (and
    sustained for some time) around the adsorbate, displaying a rich vortical dynamics.
  \end{itemize}
\end{itemize}

To conclude, we can say that, although there is no one-to-one correspondence between classical and
Bohmian trajectories, it is still possible to understand these two alternative descriptions in a
complementary way, with one being the skeleton upon which the other rests, at least at a formal level.
Classical trajectories have been and are used to understand in relatively simple terms why quantum
distributions are as they are, in a way analogous to how an optical path allows us to understand and
explain the appearance of wave phenomena, such as diffraction or interference.
Bohmian trajectories are synthesized from wave amplitudes (wave functions), so the same underlying
scheme should also be valid (i.e., using classical trajectories to understand why Bohmian
trajectories evolve in the way they do).
At the same time, Bohmian trajectories offer a clear picture of the evolution of the quantum
system by monitoring the local evolution of the quantum flux, which provides some clues on dynamical
aspects that otherwise would remain hidden (e.g., the development and effects of vortical dynamics,
or the appearance of effective barriers).

\vspace{6pt}
\acknowledgments{The author would like to express his gratitude to the Frankling-Fetzer Foundation
for support to attend the Emergent Quantum Mechanics (EmQM17) conference, held in London on the occasion
of the David Bohm Centennial, and the organizers for their kind invitation to participate in the meeting.
Support from the Spanish MINECO is also acknowledged (Grant No. FIS2016-76110-P).}


\conflictsofinterest{The author declares no conflict of interest.}


\reftitle{References}



\end{document}